\documentclass[epj]{svjour}

\usepackage{graphics}
\usepackage{lastpage}
\usepackage[utf8]{inputenc}
\usepackage{nicefrac}
\usepackage{graphicx}%
\usepackage{multirow}%
\usepackage{amssymb,amsfonts}%
\usepackage{mathrsfs}%
\usepackage[title]{appendix}%
\usepackage{xcolor}%
\usepackage{textcomp}%
\usepackage{manyfoot}%
\usepackage{booktabs}%
\usepackage{algorithm}%
\usepackage{algorithmicx}%
\usepackage{algpseudocode}%
\usepackage{listings}%
\usepackage{subcaption}
\usepackage{braket}
\usepackage{physics}
\usepackage{hyperref}
\usepackage{url}

\newcommand{\sr}{\mathrm{s}}			
\newcommand{\pr}{\mathrm{p}}			
\newcommand{\dr}{\mathrm{d}}			
\newcommand{\fr}{\mathrm{f}}			
\newcommand{\SR}{\mathrm{S}}			
\newcommand{\PR}{\mathrm{P}}			
\newcommand{\DR}{\mathrm{D}}			
\newcommand{\er}{\mathrm{e}}			
\newcommand{\ir}{\mathrm{i}}			

\begin{document}
\title{Energies and lifetimes of the 9p and 10p excited states in atomic francium}
\author{
Pierre Lassègues\inst{1,2}\thanks{\email{pierre.lassegues@cern.ch}} \and
A. Chakraborty\inst{3} \and
R.P. de Groote\inst{1} \and
B.K. Sahoo\inst{3} \and
O. Ahmad\inst{1} \and
C.M. Fajardo-Zambrano\inst{1} \and
A. McGlone\inst{4} \and
J. Warbinek\inst{5} \and
A. Ajayakumar\inst{6} \and
M. Athanasakis-Kaklamanakis\inst{7} \and
M. Au\inst{6} \and
J. Berbalk\inst{1,6} \and
D. Bettaney\inst{4} \and
B. van den Borne\inst{1} \and
T.E. Cocolios\inst{1} \and
M. Duggan\inst{4} \and
K.T. Flanagan\inst{4} \and
R.F. Garcia Ruiz\inst{8,9} \and
D. Gonzalez-Acevedo\inst{8} \and
A. Kastberg\inst{10} \and
A. Koszorús\inst{1} \and
L. Lalanne\inst{11} \and
K.M. Lynch\inst{4} \and
D.T. McElroy\inst{4} \and
G. Neyens\inst{1} \and
L. Nies\inst{5} \and
L. Quanjel\inst{1,12} \and
A. Raggio\inst{1,13} \and
J.R. Reilly\inst{4} \and
J. Snikeris\inst{14} \and
S.G. Wilkins\inst{8} \and
X.F. Yang\inst{15}
}

\institute{
KU Leuven, Instituut voor Kern-en Stralingsfysica, 3001 Leuven, Belgium \and
LNCMI, Université de Toulouse, CNRS, 31400 Toulouse, France \and
Atomic, Molecular and Optical Physics Division, Physical Research Laboratory, Ahmedabad 380009, India \and
Department of Physics and Astronomy, The University of Manchester, Manchester M13 9PL, United Kingdom \and
Experimental Physics Department, CERN, CH-1211 Geneva 23, Switzerland \and
Systems Department, CERN, CH-1211 Geneva 23, Switzerland \and
JILA and University of Colorado, Boulder, Colorado 80309, USA \and
Department of Physics, Massachusetts Institute of Technology, Cambridge, Massachusetts 02139, USA \and
Laboratory for Nuclear Science, Massachusetts Institute of Technology, Cambridge, Massachusetts 02139, USA \and
Institut de Physique de Nice, Université Côte d’Azur, 06108 Nice, France \and
Université Paris-Saclay, CNRS/IN2P3, IJCLab, 91405 Orsay, France \and
Belgian Nuclear Research Centre, SCK CEN, 2400 Mol, Belgium \and
Department of Physics, University of Gothenburg, SE-412 96 Gothenburg, Sweden \and
Institute of Atomic Physics and Spectroscopy, University of Latvia, 1004 Riga, Latvia \and
School of Physics and State Key Laboratory of Nuclear Physics and Technology, Peking University, Beijing, China
}
\abstract{
We present the first experimental determination of the $9\pr\,^2\PR_{\nicefrac{1}{2},\nicefrac{3}{2}}$ and $10\pr\,^2\PR_{\nicefrac{1}{2},\nicefrac{3}{2}}$ absolute wavenumbers, $T$, in francium, respectively as: $27111.238(5), 27360.097(5), 29058.1(10)$ and $29192.8(11)$ cm$^{-1}$, as well as their corresponding  radiative lifetimes, $\tau$: $329(6), 179(6), 553(22)$ and $362(6)$ ns. These measurements were performed using the collinear resonance ionisation spectroscopy (CRIS) experiment at ISOLDE/CERN, with a beam of $^{221}$Fr atoms. The results provide a precision experimental test of relativistic coupled-cluster theory for the heaviest alkali, showing good agreement between experiment and \emph{ab initio} calculations for lifetimes and relative excitation energies, despite a residual global offset in absolute energies.
} 
\maketitle
\emph{Acknowledgements} \hspace{1em}
The ISOLDE technical teams are gratefully acknowledged for their support. Appreciation is extended to funding and support from the following sources : KU Leuven BOF (No. C14/22/104), EOS project MANASLU (No. 40007501), FWO grants (No. G0F7321N and IRI project I001323N), EUROLABS project (No. 101057511) and STFC Grants No. ST/Y000323/1, ST/X00502X/1, and ST/V001116/1. Support from the Royal Society Grant No. DHF /R1/231007 and UK Research and Innovation (UKRI) under the UK government's Horizon Europe funding Guarantee Grant No. EP/Y036816/1 (ESPEN) and Carl Tryggers foundation for scientific research (CTS 23:2781). The work of BKS at the Physical Research Laboratory (PRL) is supported by ANRF with grant no. CRG/2023/002558 and the Department of Space, Government of India. AC and BKS acknowledge use of the ParamVikram-1000 HPC cluster at PRL, Ahmedabad, India for the RCC calculations.
\section{Introduction}
\label{intro}

The search for physics beyond the standard model is pursued at a wide range of energy scales, from the TeV range \cite{vidal2019standardmodelphysicshllhc}, to high precision experiments at $10^{-14}$ eV using molecular and atomic systems \cite{Tarbutt_2009,chupp2019,Arrowsmith-Kron_2024}. In the latter case, sought-after observables, such as the electron permanent electric-dipole moment \cite{sakemi_search_2011,Roussy2023}, parity non-conservation and the nuclear anapole moment \cite{aubin_atomic_2013,Gomez_2007} tend to scale strongly with the atomic number $Z$.
Francium,  with $Z\!=\!87$, is the heaviest alkali and therefore an attractive candidate for precision measurements of these discrete symmetry-violating observables. 
A substantial part of the francium energy spectrum remains unexplored, experimentally or theoretically with \emph{ab initio} methods. 
This leaves a problem for the analysis of precision measurements and the extraction of fundamental parameters from their data. Conclusions about fundamental physics, beyond the standard model, will necessitate reliable spectroscopic data and validated theoretical models.

\begin{figure*}[t]
	\centering
	\includegraphics[width=\textwidth]{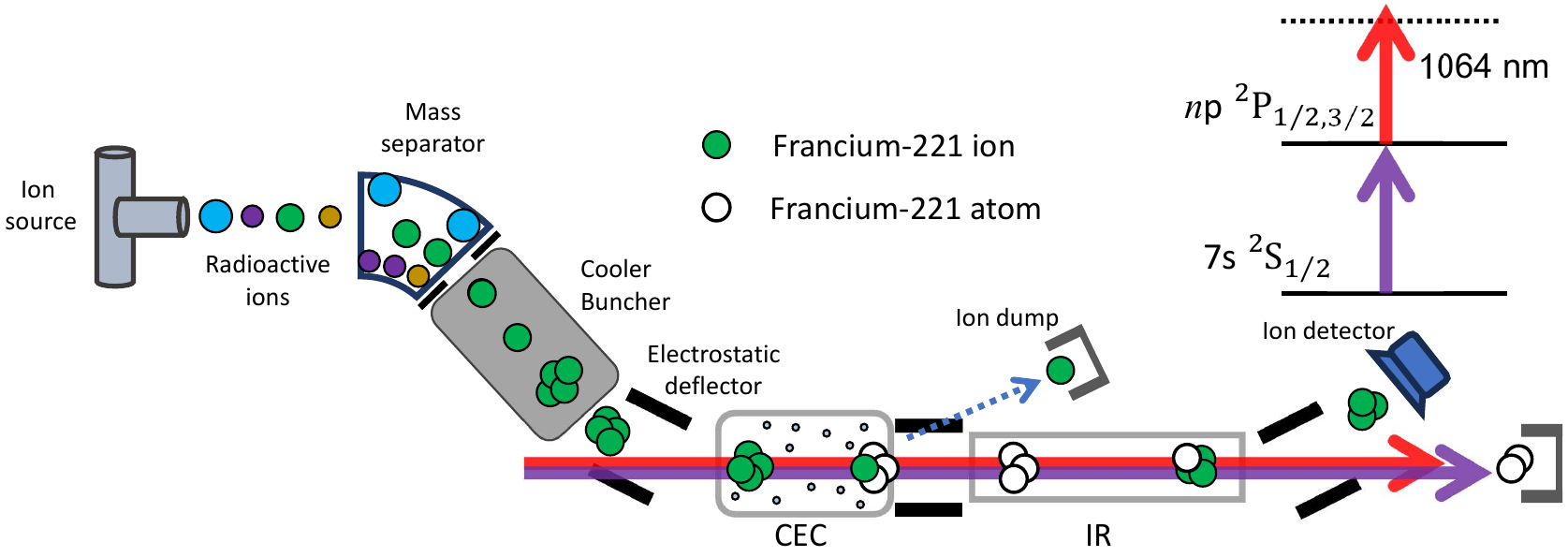}
	\caption{Overview of the CRIS beam line, detailing the sequence of ion production, mass separation, bunching within a linear Paul trap (cooler buncher), neutralisation via the charge-exchange cell (CEC), and resonant laser ionisation in the interaction region (IR). Top-right:  Example of a laser excitation sequence employed for francium in this study (shown here for excitation from ground state to a generic $n\pr\,^2\PR$ level).}
	\label{fig:CRIS}
\end{figure*} 

Accurate knowledge of the E1 (electric dipole) matrix elements is crucial for determining nuclear spin-independent and spin-dependent parity violating amplitudes \cite{aoki2018parity,kastberg2019,Gwinner_2022,PhysRevA.108.062815} and parity and time-reversal symmetry violating enhancement factors due to electron electric dipole moment \cite{CMS2021Angular} in atomic systems. The accuracy of theoretical calculations of these matrix elements can be evaluated by comparing to experimental data on lifetimes of atomic states \cite{Zhao_1997}. 

Previous theoretical studies of francium have shown that excitation energies and transition properties are strongly affected by relativistic and electron-correlation effects, with significant method-dependent corrections, particularly for highly excited states \cite{safronova_2007,PhysRevA.63.062101}. In the absence of accurate experimental benchmarks, theoretical approaches such as relativistic many-body perturbation theory and relativistic coupled-cluster (RCC) theory cannot be rigorously tested, limiting the reliable extraction of observables relevant to precision tests of fundamental symmetries \cite{safronova_2007,sahoo_correlation_2015}. 
While correlations have been studied for low angular momentum states, such as s-states \cite{Vajed_1982,Owusu_1997}, they have not been explicitly demonstrated for higher angular momentum excited states in francium, such as the p and d states.

A lot of experimental information on the level structure of francium is lacking, mostly because all francium isotopes are radioactive. The longest-lived isotope, $^{223}\mathrm{Fr}$, has a half-life of 22 minutes, while we performed studies using the shorter-lived $^{221}\mathrm{Fr}$ (half-life of 4.8 minutes) \cite{nudat3}.  This work presents measurements of the 9p and 10p absolute wavenumbers, fine-structure splittings and radiative lifetimes. These states have never been measured before in francium, due to overall difficulty of production, excitation and detection schemes.  For the 9p and 10p states, the wavenumber values published in the National Institute of Science and Technology (NIST) Atomic Spectra Database \cite{NIST-ASD}, are semi-empirical, based on Biemont \textit{et al.} \cite{biemont_theoretical_1998} and are computed using a Rayleigh-Ritz formula. 
Isotope shifts are not taken into account in that work. \textit{Ab initio} calculations of the 9p and 10p energies were reported in \cite{das_study_2019}, and radiative lifetimes were calculated with a Coulomb approximation in \cite{Weijngaarden_lifetimes_1999}.

The results presented in this paper were obtained with collinear resonance ionisation spectroscopy of the isotope \textsuperscript{221}Fr. We also present new  \textit{ab initio} calculations by employing RCC methods, performed for all experimental data presented in this work. The experimental procedure was benchmarked against previous measurements of the $8\pr\,^2\PR_{\nicefrac{3}{2}}$ state. This combined experimental and theoretical work enables  the accuracy of calculations of E1 matrix elements to be evaluated, providing essential experimental input for future atomic parity-violating amplitude measurements.\\

\section{Measurement method and experimental setup}
Our experimental data were obtained at the Collinear Resonance Ionisation Spectroscopy (CRIS) experiment using a beam of $^{221}$Fr$^+$ produced at the ISOLDE radioactive ion-beam facility at CERN \cite{kugler_isolde_2000}. The radioactive isotopes were produced upon the impact of an energetic (1.4 GeV) beam of protons from the CERN Proton Synchrotron Booster onto an uranium carbide target \cite{kugler_isolde_1993,Catherall_2017}. After irradiation, $^{221}$Fr$^+$ continued to be formed in the target, as part of the decay chain of the long-lived $^{225}$Ac- and $^{229}$Th-isotopes produced in the same reaction. By heating the pre-radiated target, francium atoms were released, and then diffused into a hot-cavity surface ionisation ion source, where they were ionised with near unity efficiency. The ions were accelerated to 30 keV, separated by their mass-to-charge ratio in the high-resolution separator magnets HRS \cite{giles_high_2003}, cooled and bunched in a gas-filled radio-frequency quadrupole linear Paul trap (ISCOOL) \cite{mane_ion_2009}. From there, a $^{221}\mathrm{Fr}$ beam with a maximum estimated intensity of $8.7\times10^{6}$ ions/s was delivered to the experimental setup (Fig.\ \ref{fig:CRIS}) \cite{Koszorus2020,Vernon2020,ATHANASAKISKAKLAMANAKIS202386}.
Entering the CRIS experiment, the ions were neutralised in a charge-exchange cell (CEC) filled with sodium vapour \cite{vernon_simulation_2019}. Non-neutralised ions were deflected to a beam dump. Neutralised francium atoms entered the interaction region where they were resonantly ionised via a two-step pulsed laser scheme (Fig.\ \ref{fig:CRIS}, top right). Following the state selective interaction with the light, generated ions were deflected onto a single-ion counter, while the residual neutral beam was discarded.

The ion beam was released from the cooler buncher with a repetition rate of 100\,Hz, in bunches of 5\,$\mathrm{\mu s}$ temporal width, which corresponds to a $0.8$\,m longitudinal spatial width. The ions were accelerated out of the cooler-buncher, held at a voltage of 29947.5(6) V, which was monitored using a calibrated voltage divider and digital multimeter \cite{lalanne_2026}.
Light for the 422 nm $7\sr\,^2\mathrm{S}_{\nicefrac{1}{2}} \rightarrow 8\pr\,^2\mathrm{P}_{\nicefrac{3}{2}}$ transition was produced by external-cavity second-harmonic generation (SHG) from a pulsed Titanium:Sapphire (Ti:Sa) laser \cite{mattolat_allsolidstate_2009}, which was pumped by a pulsed 532 nm laser at 1\,kHz.
The 365--368 nm excitation light needed for $7\sr\,^2\SR_{\nicefrac{1}{2}}{}\rightarrow{}9\pr\,^2\PR_{\nicefrac{1}{2},\nicefrac{3}{2}}$ was produced by a second Ti:Sa laser by internal cavity SHG.
The 342--343 nm excitation step for $7\sr\,^2\SR_{\nicefrac{1}{2}}{}\rightarrow{}10\pr\,^2\PR_{\nicefrac{1}{2},\nicefrac{3}{2}}$ was generated by an optical parametric oscillator (OPO) laser (Ekspla NT230), at 100 Hz repetition rate. The 1064 nm ionisation step was provided by a pulsed Nd:YAG laser at 100 Hz repetition rate for all schemes.
For lifetime studies, the excitation step was produced by the OPO for all energy levels.
All lasers were pulsed and in broadband configuration, with measured linewidths $\Delta\nu\!\sim\!3$ GHz for the two Ti:Sa, and $\sim\!200$~GHz for the OPO.
As a stable frequency reference, a diode laser was frequency stabilised using saturated absorption spectroscopy on the $5\sr\,^2\SR_{\nicefrac{1}{2}}\,F\!=\!2 \rightarrow 5\pr\,^2\PR_{\nicefrac{1}{2}}\,F'\!=\!1$ hyperfine transition in $^{87}$Rb, referenced at $12578.84851$ cm$^{-1}$ \cite{steck_rubidium_2001}. The frequency of the OPO laser was determined via an internal reference calibrated on known atomic lines in francium and rubidium. 
The Ti:Sa laser pulses at 422~nm had a full width at half maximum (FWHM) of 46(3)~ns, with a jitter of 5~ns, while the 365--368 nm Ti:Sa pulses had a FWHM of 38(2)~ns and a 4 ns jitter. The OPO laser produced pulses with a FWHM of 2.2(1)~ns and a jitter of 0.07~ns, and the Nd:YAG laser pulses had a FWHM of 10(2)~ns with 0.5 ns jitter. Laser pulse timing was controlled using an ultra-low-jitter ($<50$~ps) multi-channel signal generator.

\section{Wavenumbers of 9p and 10p configurations in neutral francium}

\begin{figure}[ht!]
    \centering
    \includegraphics[scale =0.6]{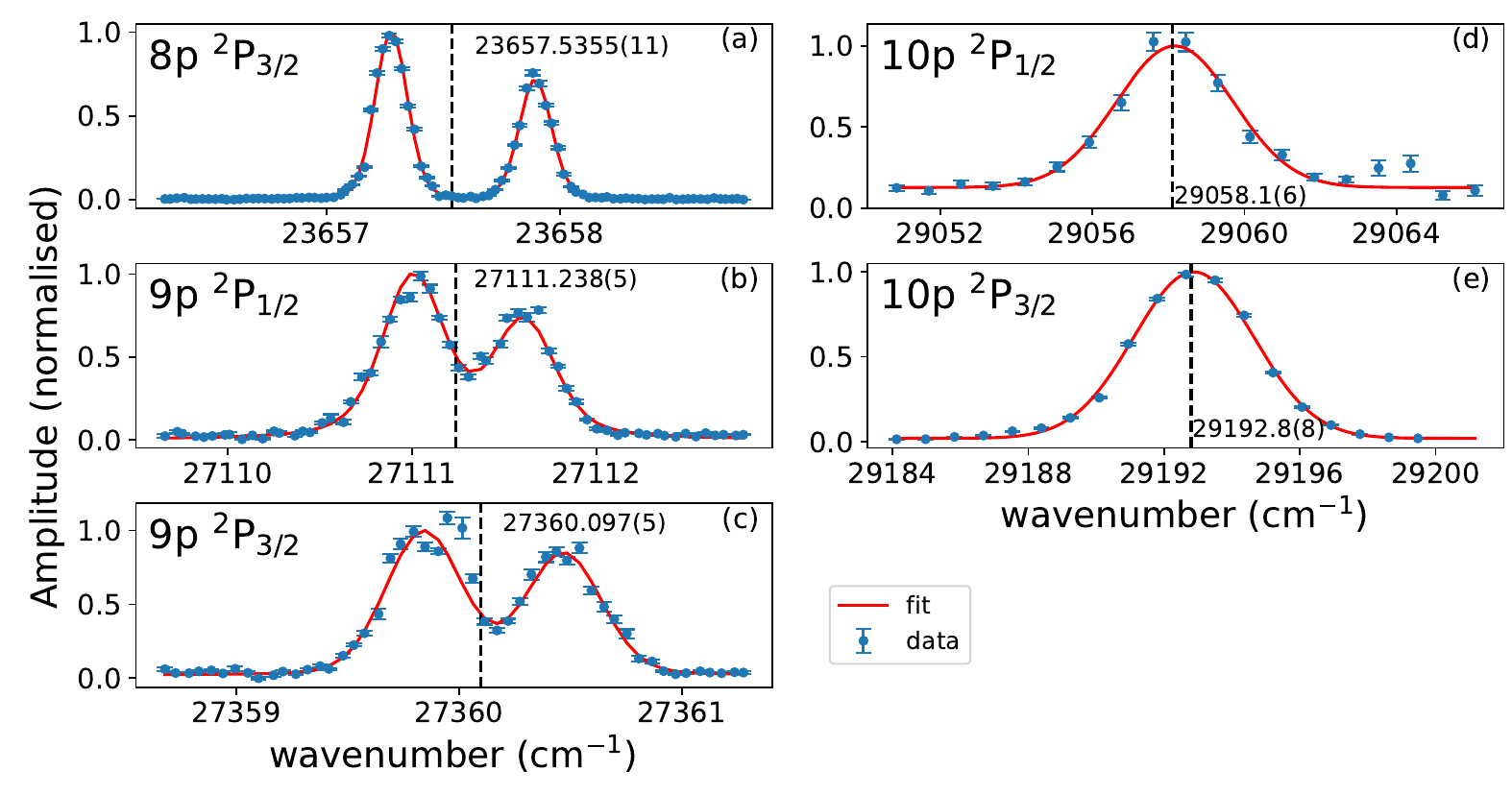}
	\caption{Measured spectra for $^{221}$Fr-transitions. On-resonance ion counting rate: 200–600 ions/s, depending on the transition. The upper-level hyperfine splittings are unresolved. For (a), (b) and (c) 
    the two-peak structure is dominated by the hyperfine splitting in the ground state $7\sr\,^2\SR_{\nicefrac{1}{2}}$ (6207 MHz \cite{Allegrini_2022,sansonetti_spectroscopic_2007}). For (d) and (e) 
     the laser linewidth was too large to be resolved. The centroid is indicated by a vertical dashed black line with the corresponding, fitted wave number given.
     }
	\label{fig:energy}
\end{figure}

Examples of resonance ionisation spectra of $^{221}$Fr are shown in Fig.~\ref{fig:energy}. The amplitude of the signal is shown as function of the rest frame wavenumber of the transition. The data were binned in wavenumber such that each spectral peak was sampled by at least 8--10 bins across its FWHM. The statistical uncertainty associated with each data point was taken as the square root of the total number of counts in each bin. For each dataset, the centroid term value and uncertainty estimates were extracted using a least-squares fit with the SATLAS package \cite{GINS2018286}.

For the reference level $8\pr\,^2\PR_{\nicefrac{3}{2}}$, the ratio of hyperfine constants 
$A(8\pr\,^2\PR_{\nicefrac{3}{2}})/A(7\sr\,^2\SR_{\nicefrac{1}{2}})=0.003608(16)$ was constrained 
\cite{Allegrini_2022}. Due to the limited laser resolution, the hyperfine structure of the upper level was not resolved, and the separation between its hyperfine components is small compared to the observed linewidth \cite{Allegrini_2022,sansonetti_spectroscopic_2007}. 
Fixing the ratio allows and using Racah intensities a reliable determination of the ground-state magnetic-dipole hyperfine coefficient, yielding $A(7\sr\,^2\SR_{\nicefrac{1}{2}}) = 6200(16)\,\mathrm{MHz}$, in agreement with the literature values obtained from measurements performed  using the same experimental setup \cite{budincevic_laser_2014}, and from compiled reference data \cite{Allegrini_2022}.

The value of $A(7\sr\,^2\SR_{\nicefrac{1}{2}})$ was held as a constant in the analyses to extract the centroids of the transitions to the $9\pr$ and $10\pr$ doublets. For these higher-lying states, the hyperfine splittings are even smaller, and the ratios between ground- and excited state hyperfine coefficients are not known. Using the ground-state A coefficient as a constraint proved necessary for stable centroid extraction. The $A$ and $B$ hyperfine coefficients of the upper states, Gaussian and Lorentzian widths as well as transition amplitudes, were kept as free parameters in the fits.

\subsection{Evaluation of experimental uncertainties in wavenumber measurements}
\label{sec: Eval}

\begin{table*}[ht]
\centering
\caption{\label{tab:frequencyuncertainties}
Summary of uncertainties in laser frequency measurements.}
\begin{tabular*}{\textwidth}{@{\extracolsep{\fill}}lll}
\toprule
\textbf{Source} & \textbf{Estimated Uncertainty} & \textbf{Notes} \\
\hline

\multicolumn{1}{c}{\textbf{\textit{Beam-Related Uncertainties}}} \\

Beam energy calibration & $\pm 4~\text{MHz}$ & Based on readout and known fluctuation ($\pm 0.6$ eV) \\

Ion beam switch & $\pm 3~\text{MHz}$ & 1\% uncertainty on a 50 V gate \\

Beam--laser angle misalignment & $-3~\text{MHz}$ & 1 cm conservative lateral offset $\rightarrow$ Doppler error \\

Charge exchange effects & Negligible for Fr & Doppler-shifted sidepeaks from excited states \cite{Vernon2019} \\

\multicolumn{1}{c}{\textbf{\textit{Laser-Related Errors}}} \\

Wavemeter accuracy & 
$\pm 10~\text{MHz}$ & Technical specification \\

Laser frequency drift & Negligible & Compared and corrected using absolute reference lines \\

OPO frequency calibration & $\pm 24~\text{GHz}$ & Calibrated using reference lines and previous measurement \\

AC Stark shift & $<1~\text{MHz}$ & Delayed high-power ionisation pulse \cite{PhysRevA.95.032502} \\

\multicolumn{1}{c}{\textbf{\textit{Other Frequency Errors}}} \\

Zeeman shift & $<1~\text{MHz}$ & Ambient field $\sim 5\text{--}10\times 10^{-5}~\text{T}$, no observed broadening \\

\bottomrule
\end{tabular*}
\end{table*}

The analyses show that the measurements of the different states are dominated by the same sources of uncertainties (see Tab. \ref{tab:frequencyuncertainties}). The statistical uncertainty is mainly governed by the laser linewidth, which limits the resolution of the hyperfine components and consequently degrades the precision of the hyperfine-structure fitting procedure, followed by a systematic contribution from the OPO laser wavenumber calibration for the 10p levels. In addition, fluctuations in the laser-induced ion rate, reflecting variations in the overall resonance ionisation efficiency and beam intensity, also contribute to this uncertainty. The following paragraphs provide a detailed discussion of the individual uncertainty contributions and the methods used to estimate them.\\
\begin{itemize}
\item  An uncertainty can be attributed to the charge exchange process, arising from the population of intermediate excited states during neutralisation. These states decay to the observed states with additional kinetic energy, causing Doppler-shifted side peaks (approximately $40~\text{MHz}$ detuned) \cite{Vernon2019}. Their contribution, especially when enhanced by inelastic collisions with sodium vapor can distort the lineshape and shift the fitted resonance centroid. However, no measurable effect attributable to this mechanism has been observed for francium \cite{de_groote_use_2015}.

\item The ion bunch is released from the ISCOOL cooler-buncher using a fast voltage pulse of approximately 50 V (slope 20 ns), which is used to open the trap every 10 ms.  This signal is used to define the extraction time and the acceleration voltage of the ions. In practice, residual potentials on the order of 1–2 V can remain during the extraction process. These residual electric fields modify the effective extraction potential experienced by the ions, leading to shot-to-shot variations in their longitudinal velocity. Such velocity variations translate into fluctuations of the Doppler shift in the collinear laser–ion interaction, which in turn increase the uncertainty of the fitted spectral line centroid.

\item To estimate the systematic frequency shift due to beam misalignment, a small angular deviation between the ion and laser beams is considered. The ion velocity is calculated classically as $v = \sqrt{2E/m}$, giving $\beta = v/c \approx 5.4 \times 10^{-4}$ and the Lorentz factor $\gamma \approx 1.00000015$. A lateral displacement of 1\,cm over a 2.36\,m propagation path, taken as a conservative estimate of the beam misalignment, corresponds to an angular deviation of $\theta \approx  4.2$\,mrad. Using the small-angle approximation $\cos(\theta) \approx 1 - \theta^2/2$, the relativistic Doppler shift for an angle $\theta$ between the laser and ion beams is approximated by:
\[
\Delta\nu \,\approx\, \nu_0  \gamma \beta \left( -\frac{\theta^2}{2} \right)  \; .
\]
For a transition wavelength of 422 nm ($\nu_0 \!=\! c/\lambda \approx 7.10 \times 10^{14}$\,Hz), the resulting systematic frequency shift is $\Delta\nu \approx -3.44$\,MHz.\\

\item The OPO laser was used for the wavenumber measurement of the 10p doublet.
In contrast to the Ti:Sa lasers used for the other levels, it could not be compared to the reference line in $^{87}$Rb, since the linewidth of the OPO laser ($\sim 200$ GHz) is too large to be read by the wavemeter. To calibrate the data, the $8\pr\,^2\PR_{\nicefrac{3}{2}}$ and $9\pr\,^2\PR_{\nicefrac{1}{2},\nicefrac{3}{2}}$ lines in $^{221}$Fr, as well as $7\pr\,^2\PR_{\nicefrac{1}{2}}$ and $8\pr\,^2\PR_{\nicefrac{1}{2},\nicefrac{3}{2}}$ in $^{87}$Rb were measured with the OPO laser. These were compared to the data taken with Ti:Sa lasers, the wavemeter and literature \cite{Morton2000_lr}. The calibrated reference frequency is determined from the fit equation 
\( y = m  x \!+\! k = 0.999707(1)\,x - 1.47(2) \). 
We propagate the uncertainties from \( m \), \( k \), and the measured value \( x \) give an uncertainty of \( 0.79~\text{cm}^{-1} \), equivalent to approximately \( 24~\text{GHz} \).
\end{itemize}

\subsection{RCC calculations}

The experimental data  is compared to theoretical calculations obtained using the RCC method with the singles, doubles and triples excitation approximation (RCCSDT method).\cite{hughes_coupled-cluster_1993,pal_relativistic_2007}. This framework provides a consistent treatment of electron correlations. The experiment measured the absolute energy difference between the $7\sr\,{}^2\SR_{1/2}$ ground state and an excited state, whereas the \textit{ab initio} calculations provide valence-electron binding (removal) energies referenced to the ionisation potential. To enable a direct comparison, the $7\sr\,{}^2\SR_{1/2}$ ground-state binding energy was calculated, yielding 23\,811.67(13)\,cm$^{-1}$, which differs from the value reported in \cite{NIST-ASD} by 154 cm$^{-1}$. Contributions to this discrepancy include missing higher-order correlation effects, including quadruple excitations and contributions from high-angular-momentum orbitals, which require more computational capacity than currently available. 
In contrast, such contribution to 8p, 9p and 10p are significantly smaller, and will remain within overall uncertainty bars. Therefore, a more robust way to compare the calculated binding energy with the experimental excitation wavenumber is to also compare the latter to the ionisation potential. Thus, 
to facilitate direct comparison with experimental excitation energies, the 7s energy was adjusted so that the theoretical 7s$\rightarrow$8p transition matches the experimental value, and all other excitation energies are then reported relative to this adjusted ground state. The remaining discrepancies, which scatter less than 5 cm$^{-1}$, can then be explained by smaller effects not fully captured in the present treatment, such as residual higher-order correlations, finite-basis limitations, or quantum electrodynamics (QED) contributions.\\

To understand the results theoretically, the measured observables have been calculated using RCC theory \cite{Sahoo_relativistic_2018,yamanaka_probing_2017}. An atomic wave function of an atomic state with closed-core $[6\pr^6]$ and a valence orbital $v$ of francium in the RCC theory framework can be expressed as 
\begin{equation}
    \ket{\Psi_\mathrm{v} } \,=\,  \er^T \Bqty{1+S_\mathrm{v} } \ket{ \Phi_\mathrm{v}} \; ,
    \label{eqcc}
\end{equation}
where $|\ket{ \Phi_\mathrm{v}}$ is the Dirac-Hartree-Fock (DHF) wave function, $T$ is the RCC operator that accounts for the excitations of core electrons to the virtual orbitals, and $S_\mathrm{v}$ is responsible for exciting electron from the valence orbital correlating with core electrons to the virtual orbitals. For accurate calculations of the DHF wave functions of all the considered states of Fr, we have employed an extensive size basis-set of primitive Gaussian-type orbitals (GTOs). The basis-set consisted of 40, 39, 38, 37, 36, 35, and 34 GTOs for the $\sr, \pr, \dr, \fr, \mathrm{g}, \mathrm{h},$ and $\ir$ symmetries,
respectively.

The amplitude and energy ($E_\mathrm{v}$) determining equations of the RCC theory are given by
\begin{equation}
    \mel{\Phi^*}{(\overline{H}_\mathrm{N} \!-\! E_\mathrm{v}) S_\mathrm{v} + \overline{H}_\mathrm{N}}{\Phi_\mathrm{v}} \,=\, 0  
\end{equation}
and 
\begin{equation}
    E_v \,=\, \ev{\overline{H}_\mathrm{N} S_\mathrm{v} + \overline{H}_\mathrm{N}}{\Phi_\mathrm{v}}
\end{equation}
respectively. In the above expressions, $\ket{\Phi^*}$ denotes an excited state determinant with respect to $\ket{\Phi_\mathrm{v} }$ and appropriately selected to evaluate amplitudes of the $T$ and $S_\mathrm{v}$ operators. $H_\mathrm{N}$ is the normal-ordered atomic Hamiltonian. For convenience, this is defined with respect to the $[6\pr^6]$ configuration, so that $E_\mathrm{v}$ directly corresponds to the negative of the ionisation potential of the valence electron. Calculations are initiated with the Dirac-Coulomb approximation in the Hamiltonian, then corrections from the Breit and QED effects are added separately. Since both amplitude and $E_\mathrm{v}$ equations are interdependent, the above equations are solved simultaneously.

After obtaining wave functions of atomic states, matrix elements of the electric dipole (E1) operator, $D$, between two states described by wave functions $\ket{\Psi_\fr }$ and $\ket{\Psi_\ir }$, which is needed for theoretical evaluations of radiatiove lifetimes, are evaluated as
\begin{equation}
    \mel{\Psi_\fr}{D}{\Psi_\ir} \,=\, \frac{ \mel{\Phi_\fr}{\{1+ S_\fr^{\dagger}\} \bar{D} \{1+S_\ir\}}{\Phi_\ir} }{\sqrt{N_\ir N_\fr\,}}  \; , 
\end{equation}
where $ N_{\mathrm{v}=\ir,\fr} = \ev{\{1+S_\mathrm{v}^{\dagger}\} \er^{T^{\dagger}} \er^T \{1+S_\mathrm{v}\}}{\Phi_\mathrm{v}}$ and $\bar{D} = \er^{T^{\dagger}} \!D\, \er^T$. The non-terminating series appearing in these expressions are accounted for by computing them through self-consistent procedures.

\subsection{Discussion on the experimental and theoretical term values}

For the theoretical values, the different uncertainty estimates reported in the literature should be interpreted with caution, as they do not necessarily refer to the same kind of accuracy. The few $0.01~\mathrm{cm^{-1}}$ tabulated in \cite{NIST-ASD} are from 
\cite{biemont_theoretical_1998}, and are associated with spectroscopic treatments and extrapolation of the quantum-defect curves. Therefore, these benefit from experimental information, and moreover that work does not include isotope shifts. In contrast, the $\sim0.1\%$ accuracy reported for the Brueckner energies of \cite{PhysRevA.63.062101} reflects the accuracy of an \emph{ab initio} many-body calculation, without such level-by-level spectroscopic extrapolation. RCC calculations, which may achieve relative uncertainties of order $\sim 0.005\%$, employ a different treatment of electron correlation and can therefore provide substantially improved theoretical accuracy. Consequently, these uncertainty estimates should not be directly compared as simple measures of the relative quality of the respective methods.

\begin{figure}[ht]
    \centering

    \begin{minipage}[c]{0.45\textwidth}
        \centering
        \includegraphics[width=\textwidth]{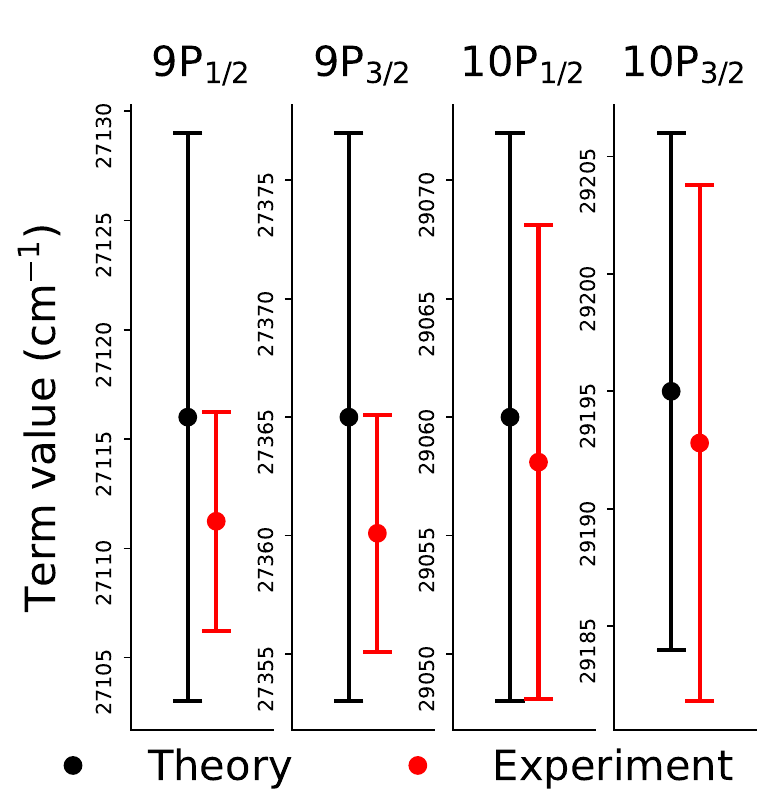}
        \captionof{figure}{Comparison of the experimental and RCCSDT term values. Vertical error bars represent the corresponding theoretical and experimental uncertainties.
}
        \label{fig:energiescomp}
    \end{minipage}
    \hspace{0mm}
    \begin{minipage}[c]{0.50\textwidth}
        \centering
        \captionof{table}{Measured and calculated absolute wavenumbers for $^{221}$Fr.}
        \label{tab:tablefreq}

        \begin{tabular}{@{}lcc@{}}
            \toprule
            Level & \multicolumn{2}{c}{Wavenumber, $T$ (cm$^{-1}$)} \\
            \cmidrule{2-3}
            & Theory & Experiment \\
            \midrule

            $8\pr\,^2\PR_{\nicefrac{3}{2}}$
            & $23661(24)^\mathrm{d}$
            & $23657.5355(11)^\mathrm{a}$ \\
            & 
            & $23657.529(5)^\mathrm{c}$ \\[1mm]

            $9\pr\,^2\PR_{\nicefrac{1}{2}}$
            & $27116(13)^\mathrm{a}$
            & $27111.238(5)^\mathrm{a}$ \\
            & $27118.21(5)^\mathrm{b}$ & \\
            & $27104(27)^\mathrm{d}$ & \\[1mm]

            $9\pr\,^2\PR_{\nicefrac{3}{2}}$
            & $27365(12)^\mathrm{a}$
            & $27360.097(5)^\mathrm{a}$ \\
            & $27366.20(5)^\mathrm{b}$ & \\
            & $27355(27)^\mathrm{d}$ & \\[1mm]

            $10\pr\,^2\PR_{\nicefrac{1}{2}}$
            & $29060(12)^\mathrm{a}$
            & $29058.1(6)[8]^\mathrm{a}$ \\
            & $29064.18(5)^\mathrm{b}$ & \\[1mm]

            $10\pr\,^2\PR_{\nicefrac{3}{2}}$
            & $29195(11)^\mathrm{a}$
            & $29192.8(8)[8]^\mathrm{a}$ \\
            & $29198.09(5)^\mathrm{b}$ & \\

            \bottomrule
        \end{tabular}

        \vspace{2mm}

        \footnotesize
        Refs. $^\mathrm{a}$This work;
        $^\mathrm{b}$\cite{biemont_theoretical_1998} uncertainty estimated from the article data, and compensated for isotope shift;
        $^\mathrm{c}$Data from \cite{duong_first_1987} corrected with an estimated isotope shift;
        $^\mathrm{d}$\cite{PhysRevA.63.062101}.
        ( ) statistical; [ ] systematic uncertainties, given when significant,
        see Tab.~\ref{tab:frequencyuncertainties}.
    \end{minipage}

\end{figure}

The comparison between the theoretical and experimental term values reveals consistent trends for the 9p and 10p configurations of francium (Fig.\ \ref{fig:energiescomp}). 

The fine-structure splitting 
within the P-terms 
decreases from approximately $249~\mathrm{cm^{-1}}$ for the 9p configuration to $135~\mathrm{cm^{-1}}$ for 10p.
This trend is consistent with the increasing spatial extent of the valence electron wavefunction with increasing $n$, which reduces its overlap with the inner region of the atomic potential, where relativistic and spin-orbit effects are strongest.
 
The comparison with the RCCSDT results shows excellent agreement, with residual discrepancies, below $5~\mathrm{cm^{-1}}$ for all four levels. After adjusting the theoretical 7s wavenumber to reproduce the experimental $7\mathrm{s}\!\rightarrow8\!\mathrm{p}$ transition, these small remaining differences can be attributed to residual higher-order correlation effects, finite-basis limitations, and possible QED contributions. The observed agreement therefore indicates that the RCCSDT approach provides an accurate description of the relativistic electronic structure and fine-structure splittings of the excited p configurations of francium.

A summary of the results is shown in Tab.\ \ref{tab:tablefreq}. The centroid of the $7\sr\,{}^2\SR_{\nicefrac{1}{2}} \rightarrow 8\pr\,{}^2\PR_{\nicefrac{3}{2}}$ transition for $^{212}$Fr and isotope shift between $^{212}$Fr and $^{221}$Fr have been previously measured, corresponding to a transition wavenumber of $23\,657.529(5)\ \mathrm{cm}^{-1}$ for $^{221}$Fr ~\cite{duong_first_1987,sansonetti_spectroscopic_2007}. Our value of 23657.5355(11) cm$^{-1}$ is in agreement with this literature value within uncertainties.
\section{Radiative lifetimes of $n$p configurations}

\begin{figure}[htbp]
    \centering

    \begin{subfigure}[b]{0.48\textwidth}
        \centering
        \includegraphics[width=\linewidth]{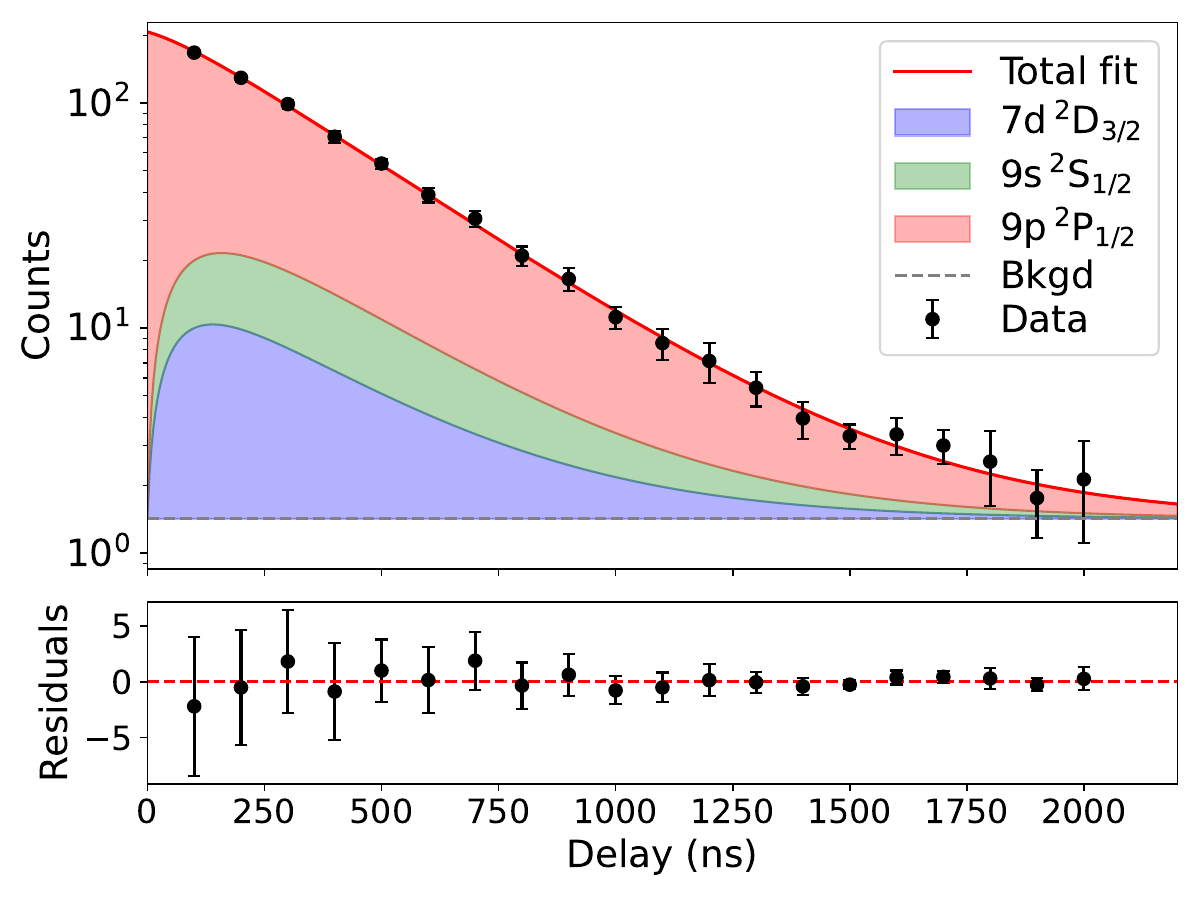}
        \caption{}
        \label{fig:a}
    \end{subfigure}
    \hfill
    \begin{subfigure}[b]{0.48\textwidth}
        \centering
        \includegraphics[width=\linewidth]{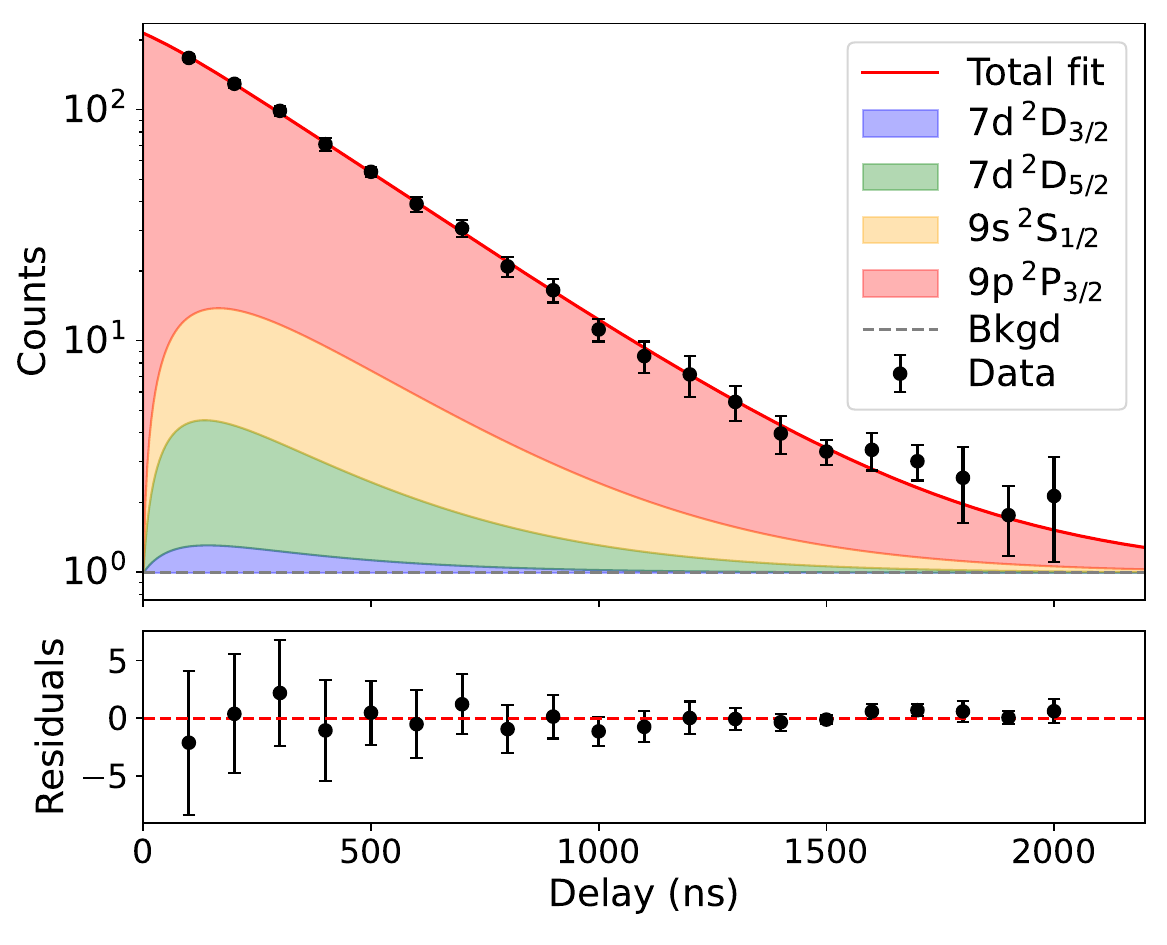}
        \caption{}
        \label{fig:b}
    \end{subfigure}

    \vspace{0.3cm}

    \begin{subfigure}[b]{0.48\textwidth}
        \centering
        \includegraphics[width=\linewidth]{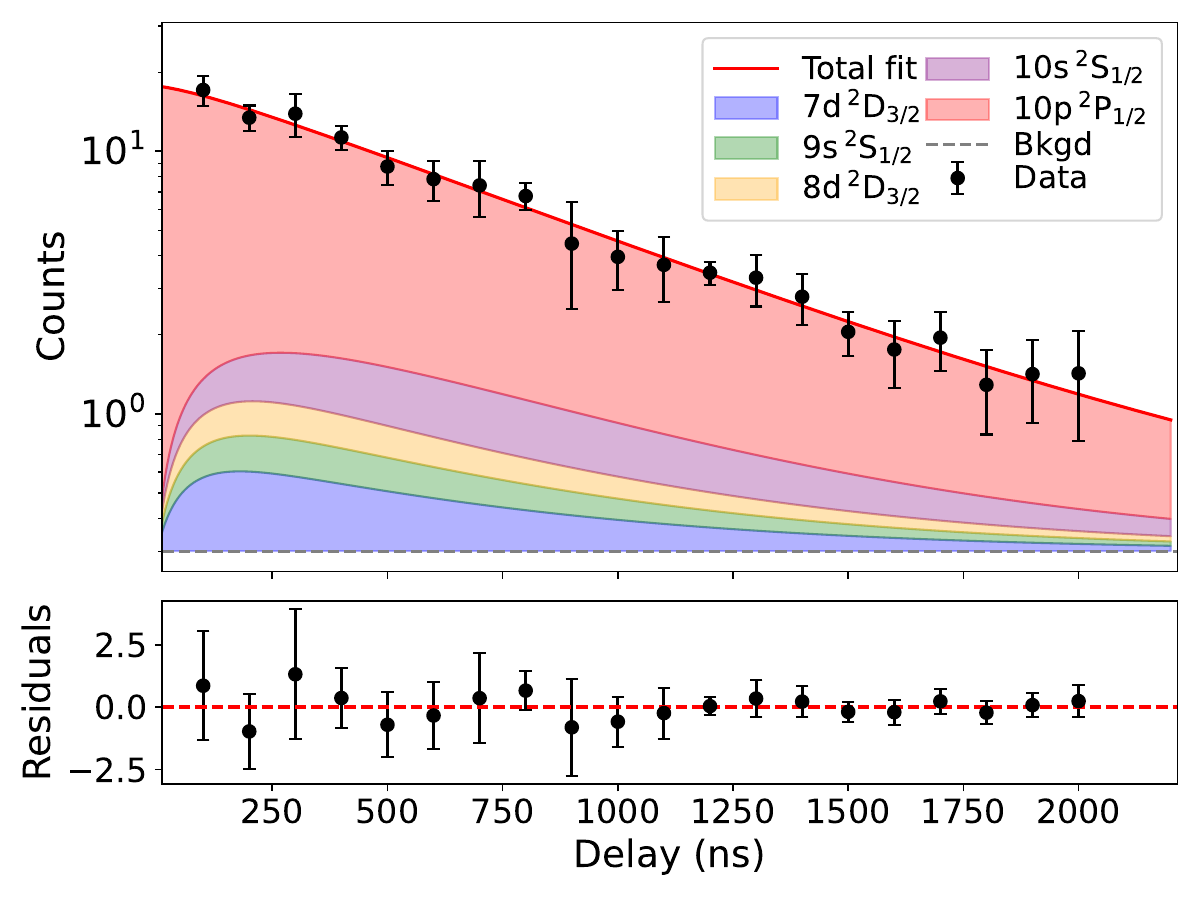}
        \caption{}
        \label{fig:c}
    \end{subfigure}
    \hfill
    \begin{subfigure}[b]{0.48\textwidth}
        \centering
        \includegraphics[width=\linewidth]{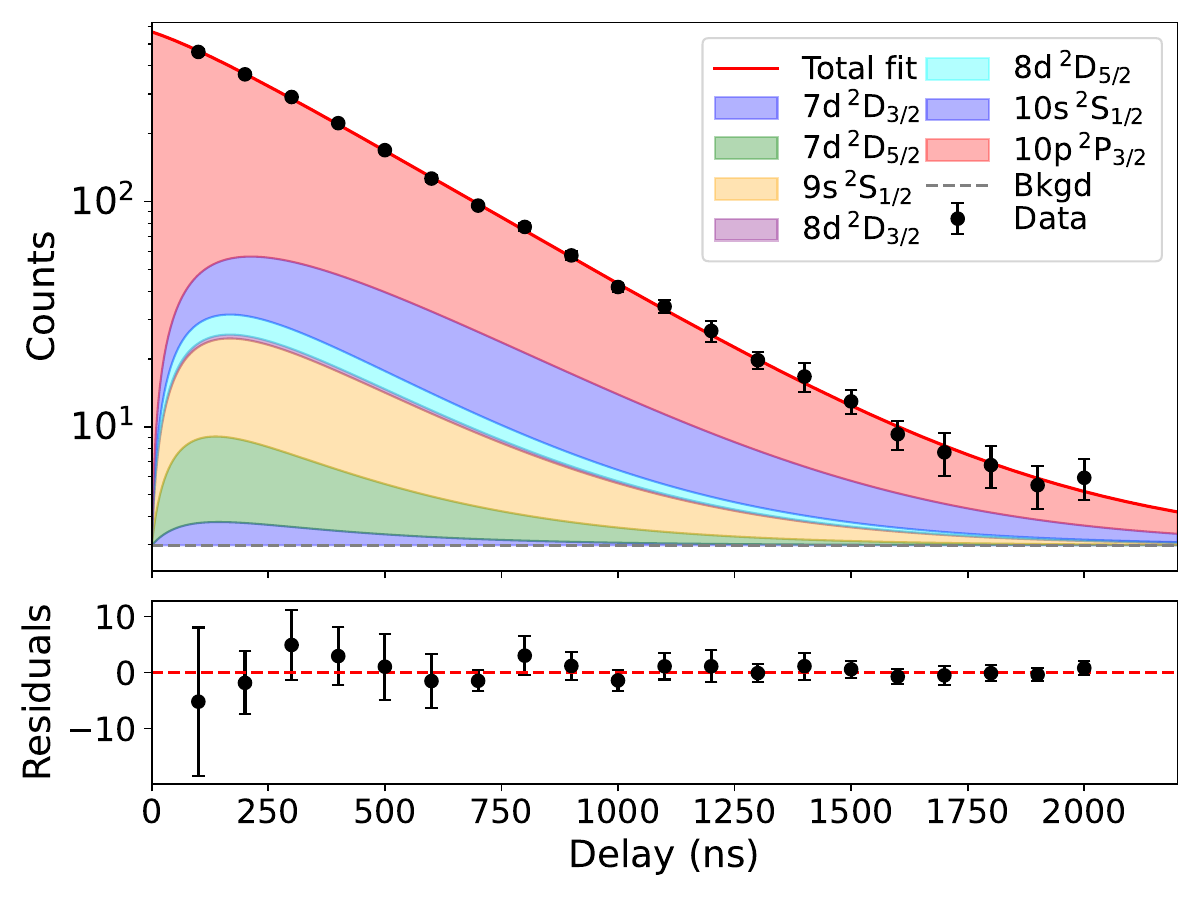}
        \caption{}
        \label{fig:d}
    \end{subfigure}

    \caption{Decay curves for (a) $9\pr\,^2\PR_{\nicefrac{1}{2}}$, (b) $9\pr\,^2\PR_{\nicefrac{3}{2}}$, (c) $10\pr\,^2\PR_{\nicefrac{1}{2}}$, (d) $10\pr\,^2\PR_{\nicefrac{3}{2}}$ level, with contribution of lower-lying states to the total population. The vertical axis uncertainties account for both statistics and systematics. The contributions of the ionised cascade levels are shown with different colours. Bkgd : Background}
    \label{fig:panel22}
\label{fig:lifetimes}
\end{figure}

\begin{table}[h]
\caption{\label{tab:tableltallP}%
Details of the different cycles in the
lifetime measurement. $8\pr\,^2\PR_{\nicefrac{3}{2}}$ cycles 1--4 were previously reported in \cite{athanasakis-kaklamanakis_radiative_2024}. The systematic
signal uncertainties correspond to the added uncertainties on the count rate, in order 
to account for atomic beam fluctuations.}

\centering
\begin{tabular}{@{}lcccc@{}}
\toprule
Level  &&  ISCOOL  &  Syst.\  &  Lifetime, $\tau$  \\
&  n°  &  ejection  &  signal &  (ns)  \\
&&  delay($\mu$s)  & uncert.(\%)  \\ 
\hline
$8\pr\,^2\PR_{\nicefrac{3}{2}}$  &  1  &  0  &  3.19  &  80.97(6.28)  \\
&  2  &  $-0.5$  &  3.1  &  77.68(7.23)  \\
&  3  &  $+0.5$  &  1.09  &  101.03(6.87)  \\
&  4  &  0  &  5.87  &  77.97(4.79)  \\
\hline
&  5  &  0  &  7.84  &  71.68(6.41)  \\
&  6  &  $+1$  &  1.18  &  81.71(3.44)  \\
&  7  & $-1$  &  1.31  &  80.92(2.86)  \\
&&&weighted mean:  &  \textbf{81.14(1.71)}  \\
\hline
ref.\ \cite{aubin_lifetimes_2004}  &&&&  83.5(1.5)  \\
\hline
\hline
$9\pr\,^2\mathrm{P}_{1/2}$  &  1  &  0  &  1.94  &  320.45(10.93)  \\
&  2  &  $+1$  &  1.62  &  344.94(8.92)  \\
&  3  &  $-1$  &  1.09  &  318.42(9.50)  \\
&&&weighted mean:  & \textbf{329.36(5.59)}  \\
\hline
$9\pr\,^2\mathrm{P}_{3/2}$  &  1  &  0  &  0.98  &  175.66(9.52)  \\
&  2  &  $+1$  &  1.15  &  183.73(6.65)  \\
&  3  &  $-1$  &  0.99  &  174.74(7.36)  \\
&&&weighted mean:  & \textbf{178.84(4.38)}  \\
\hline

$10\pr\,^2\mathrm{P}_{1/2}$  &  1  &  0  &  1.99  &  565.29(30.87)  \\
&  2  &  $+1$  &  2.58  &  570.10(55.57)  \\
&  3  &  $-1$  &  1.91  &  535.52(31.46)  \\
&&&weighted mean:  & \textbf{553.32(20.48)}  \\
\hline
$10\pr\,^2\mathrm{P}_{3/2}$  &  1  &  0  &  0.49  &  358.39(7.54)  \\
&  2  &  $+1$  &  0.55  &  383.14(8.91)  \\
&  3  &  $-1$  &  0.36  &  342.96(9.69)  \\
&&&weighted mean:  &  \textbf{362.00(4.95)}  \\

\bottomrule
\end{tabular}
\end{table}

Having established the comparison between theoretical and experimental term values, we now turn to radiative lifetimes, which provides a test of the accuracy of the calculated electric-dipole (E1) matrix elements. All lifetime measurements were performed using a two-step laser scheme, as previously described. First, using the OPO laser to excite the atom to the desired state, and then the Nd:YAG laser to ionise the atom. To determine the lifetimes of the different levels, we measured the average count rate on the ion detector over a 120 second period, for several different delays between the two laser pulses. The optimal ISCOOL trap ejection timing was determined by maximising the signal rate, thus optimising the spatial and temporal overlap between the atomic bunch and the laser pulses. Lifetime measurements were taken for the optimal ejection timing, and $\pm 1$ $\mathrm{\mu}$s compared to optimal, to constrain potential systematic effects related to the ejection timing. The lifetime measurements were distributed over two experimental campaigns. From the combined data, weighted means and uncertainty limits are derived. This is further accounted for in Tab.\ \ref{tab:tableltallP}.

For each data point, the statistical uncertainty was taken as the square root of the total number of ions detected over the measurement period. The systematic uncertainty on the signal rate was estimated by taking reference points at zero delay between the first and second excitation step every five measurements, to monitor fluctuations in the atomic beam and laser parameters. Time scale uncertainties for delays are the pulse-to-pulse timing instability (jitter) of the lasers, determined as 0.5 ns.

The excited level does not only decay to the ground state but also to intermediate lower-lying states, predominantly via allowed E1 transitions. As the 1064 nm laser can ionise states above 23450 cm$^{-1}$, the total measured signal is the sum of the population of the initial excited $n$p-level, the contributions of states in the radiative decay cascade, and a constant background
\[
N_\text{total}(t) \,=\, N_a(t) + \sum_j N_j(t) + C
  \;  ,\]
where $N_a(t)$ is the population of the initial level, $N_j(t)$ are the cascade contributions and $C$ is the background. 

The upper-level population decays exponentially
\begin{equation}
    N_a(t) \,=\, N_0\, \mathrm{e}^{-t/\tau_a}   \;  ,
\end{equation}
while the contribution of the lower-lying levels $j$ is modelled as
\begin{equation}
N_j(t) \,=\, N_0 \, b_{aj} \, \frac{\tau_j}{\tau_a - \tau_j} 
\pqty{\mathrm{e}^{-t/\tau_a} - \mathrm{e}^{-t/\tau_j}}   \;  ,
\label{eq:Nj}
\end{equation}
where $\tau_a$ and $\tau_j$ are the lifetimes of the upper and lower levels, respectively, $N_0$ the initial population, and $b_{aj}$ is the branching ratio from $a$ to $j$. In doing so, it is assumed that the ionisation cross section from all atomic states is the same. Only direct feeding of state $j$ from the $n$p state of interest is considered, and feeding of $j$ from another state $j'$ is neglected, as these feedings are strongly suppressed by the branching ratios, which favour decays to states below the ionisation threshold. Their inclusion does not affect the extracted upper-level lifetime within the experimental uncertainties. The inclusion of cascade contributions results in a modest but consistent improvement in fit quality, diminishing the reduced $\chi^2$ by approximately 0.15 on average, compared to a simple exponential decay model. The branching ratio is obtained from the transition probability of the specific decay channel relative to the total transition probability, which is determined from the lifetime of the excited state. (see Tab.\ \ref{Tab-BR}, right part). Note that this cascade model does not add any additional fit parameters compared to a simple exponential fit.

The background level was independently recorded prior to the lifetime measurement. In the fitting procedure, the background contribution \textit{C} was treated as a free parameter constrained within the range defined by the measured background mean ± one standard deviation. This approach incorporates the experimental uncertainty of the background while excluding unphysical values. 

The uncertainty was taken as the standard error 
$\sigma = \left( \sum_i 1/\sigma_i^2 \right)^{-1/2}$, 
increased by $\sqrt{\chi^2_\nu}$ when the reduced chi-squared exceeded unity. 
This adjustment avoids underestimation of uncertainties while preventing unnecessary deflation when the data are consistent. 
In addition, the uncertainties associated with the cascade parameters, namely the lifetimes $\tau_j$ in Eq.~\ref{eq:Nj}, were propagated through the full model and included as a systematic contribution.  
To exclude data points where the laser pulses overlap, which would lead to deviations from the simple decay law, the initial 15 ns of the decay curve were omitted from the fits. This interval, referred to as the ``laser overlap regime'' in previous work \cite{athanasakis-kaklamanakis_radiative_2024}, corresponds approximately to the sum of the lasers FWHM and their relative timing jitter. Typical curves are shown in Fig.\ \ref{fig:lifetimes}; no systematic deviations are seen for small decay times, confirming the validity of the chosen delay timings.

\begin{figure}[t]
\centering
\includegraphics[scale =0.6]{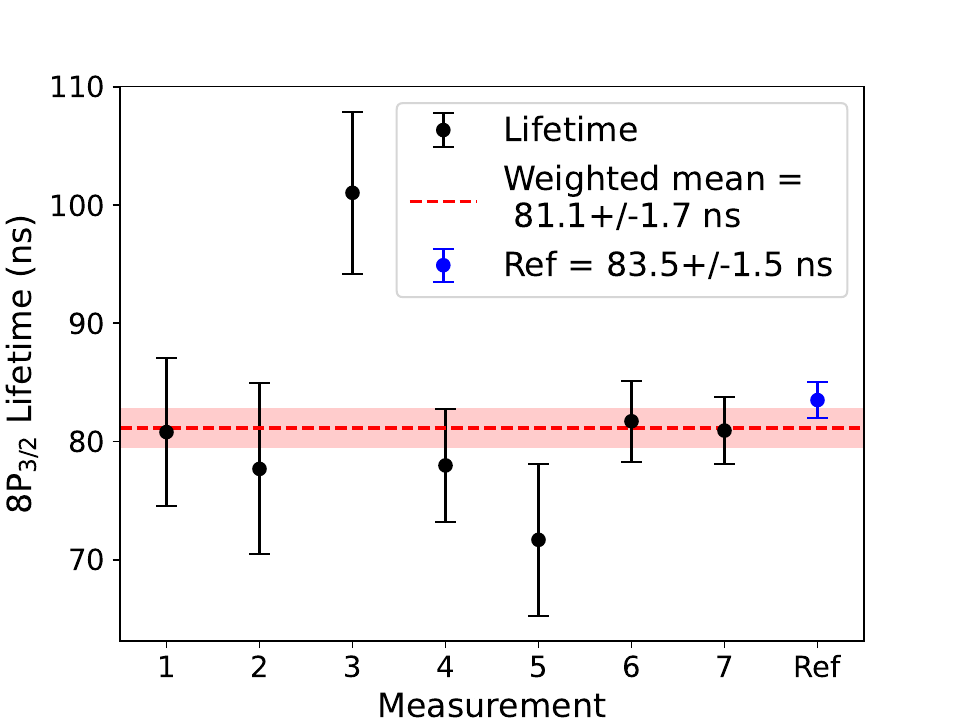}
\caption{Lifetime results for the $8\pr\,^2\PR_{\nicefrac{3}{2}}$ level. The red line and shaded band indicate the error-weighted mean and one standard deviation calculated across all measurements, (ref.\ from \cite{aubin_lifetimes_2004}).}
\label{fig:lifetime8P}
\end{figure}

To assess the reliability of the experimental technique and analysis method, the lifetime of the $8\pr\,^2\PR_{\nicefrac{3}{2}}$ level was measured using the same procedure and compared to the value reported in the literature \cite{aubin_lifetimes_2004}. Measurements 1--4 were recorded in the 2023 experimental campaign, where they also served as reference for RaF lifetime measurements (details in \cite{athanasakis-kaklamanakis_radiative_2024}). Measurements 5--7 are from the 2024 campaign. The results, presented in Fig.\ \ref{fig:lifetime8P}, give a weighted mean of $81.1(1.7)$ ns, which is consistent with the literature value of 83.5(1.5) ns \cite{aubin_lifetimes_2004}. Measured lifetimes are compiled in Tab.\ \ref{tab:tableltallP}.

\subsection{Evaluation of experimental uncertainties in lifetime measurements}

Sources of experimental uncertainties for the lifetime measurements have been listed in Tab.\ \ref{tab:lifetimeuncertainties}. The dominant contribution to the systematic uncertainty arises from the lifetimes of lower-lying cascade levels, $\tau_j$, the value of which were taken from published experimental or theoretical data, including uncertainties (Tab. \ref{Tab-BR}). Because the initially excited level decays not only to the ground state but also populates these intermediate states according to the branching ratios $b_{aj}$, the measured ion signal includes both the direct decay of the upper level, $N_a(t)$, and the cascade contributions, $N_j(t)$, as in Eq.~\eqref{eq:Nj}. Uncertainties in the $\tau_j$ values propagate into the total signal and thereby affect the fitted upper-level lifetime $\tau_a$. This effect is accounted for by individually  varying each $\tau_j$ within its reported uncertainty and refitting the decay curves; the resulting spread in $\tau_a$ is treated as a systematic contribution. 

\begin{table*}[ht!!]
\centering
\caption{\label{tab:lifetimeuncertainties}
Summary of uncertainties in lifetime measurements. * dataset-specific}
\begin{tabular*}{\textwidth}{@{\extracolsep{\fill}}lll}
\toprule
\textbf{Source} & \textbf{Estimated Uncertainty} & \textbf{Notes} \\
\hline

Cascade level lifetimes & $\pm 2$--$8~\text{ns}$ * & Lower-state uncertainties propagated through cascade fit \\

Laser timing jitter &
\begin{tabular}{@{}l@{}}
$\pm 0.07~\text{ns}$ (OPO) \\
$\pm 0.49~\text{ns}$ (Nd:YAG)
\end{tabular}
& Std.\ dev.\ of pulse centroids (FWHM: 2.2 ns, 10 ns) \\

$t_0$ offset (Laser--DAQ sync) & $<1~\text{ns}$ & Extracted from timing reference pulses \\

Beam intensity fluctuations & $\pm 0.5$--$5\%$ * & Zero-delay references to account for beam fluctuations \\

Laser-ion signal fluctuations & $\pm 1$--$15\%$ & Statistical uncertainty from integrated counts \\

Experimental background & $\pm 0.3$--$4\%$ * & Extracted from gate-closed, laser-off, and laser-detuned data \\

\bottomrule
\end{tabular*}
\end{table*}

In addition, fluctuations in the laser-induced ion rate, which reflect variations in the overall Resonance Ionization Spectroscopy (RIS) efficiency and the incoming ion flux, contribute further to the total uncertainty. The RIS efficiency itself depends on several experimental factors, including the laser ionisation cross-section, the neutralisation efficiency of the atomic beam, transport losses in the neutral and ion beams, and the ion detector efficiency. Variations in any of these parameters lead to changes in the measured signal counts, thereby affecting the decay curve data and ultimately the precision of the fitted lifetimes.
The following paragraphs provide a details on the individual uncertainty contributions and the methods used to estimate them.

\begin{itemize}

\item Laser pulses have been characterised before the experiment. The timing characteristics of each laser (OPO and Nd:YAG) used for the lifetime measurements were recorded using a Lecroy WavePro 7 Zi oscilloscope and two silicon photodiode detectors, to extract both the pulse shapes and their shot-to-shot stability (jitter). These detectors, each aligned to monitor a separate laser beam, were positioned on the laser table ahead of the beamline, with signal levels kept below saturation. The oscilloscope was externally triggered by a low-jitter ($<\! 50$ ps) Quantum Composers 9520 Pulse Generator. To characterise the pulse shape, 1000 consecutive pulses were recorded for both lasers. The resulting waveforms were averaged, providing timing profiles. These profiles (Fig.\ \ref{fig:laserpulse}) were then analysed using a skewed Voigt profile, to extract the full-width at half-maximum (FWHM), 2.2 ns for the OPO and 10 ns for the Nd:YAG at the centroid. The timing jitter was determined as the standard deviation of the centroid distribution, with a value of 0.07 ns for the OPO and 0.49 ns for the Nd:YAG.

\begin{figure*}[h!]
    \centering
    \begin{subfigure}[t]{0.45\textwidth}
        \centering
        \includegraphics[width=\linewidth]{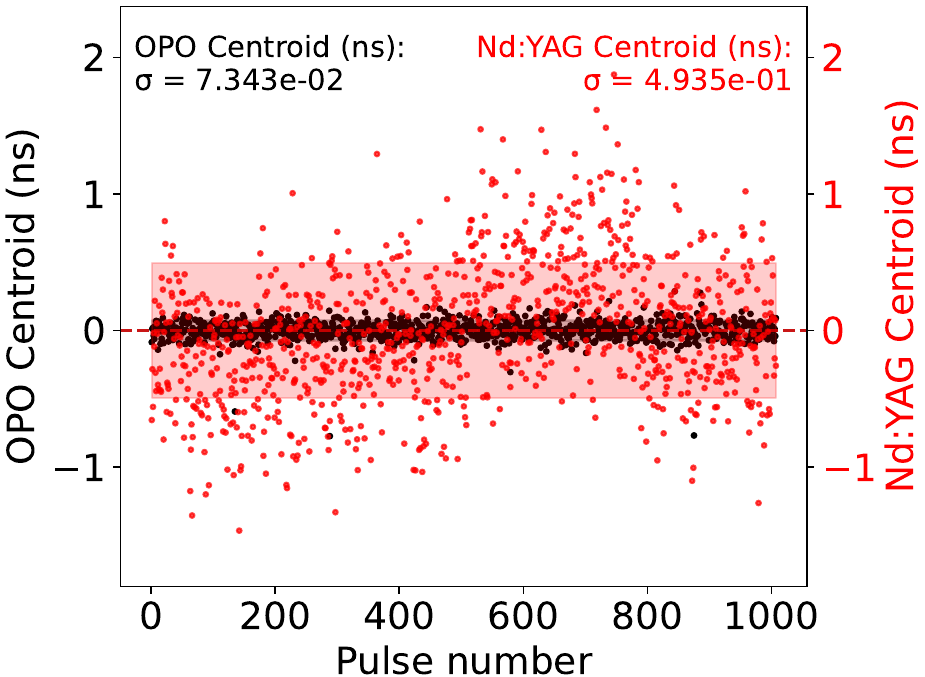}
        \caption{Shot-to-shot fluctuations: Centroid positions of each pulse relative to the mean, with standard deviation shown for both lasers (surrounding band). OPO laser (black) and Nd:YAG laser (red).}
    \end{subfigure}
    \hspace{5mm}
    \begin{subfigure}[t]{0.48\textwidth}
        \centering
        \includegraphics[width=\linewidth]{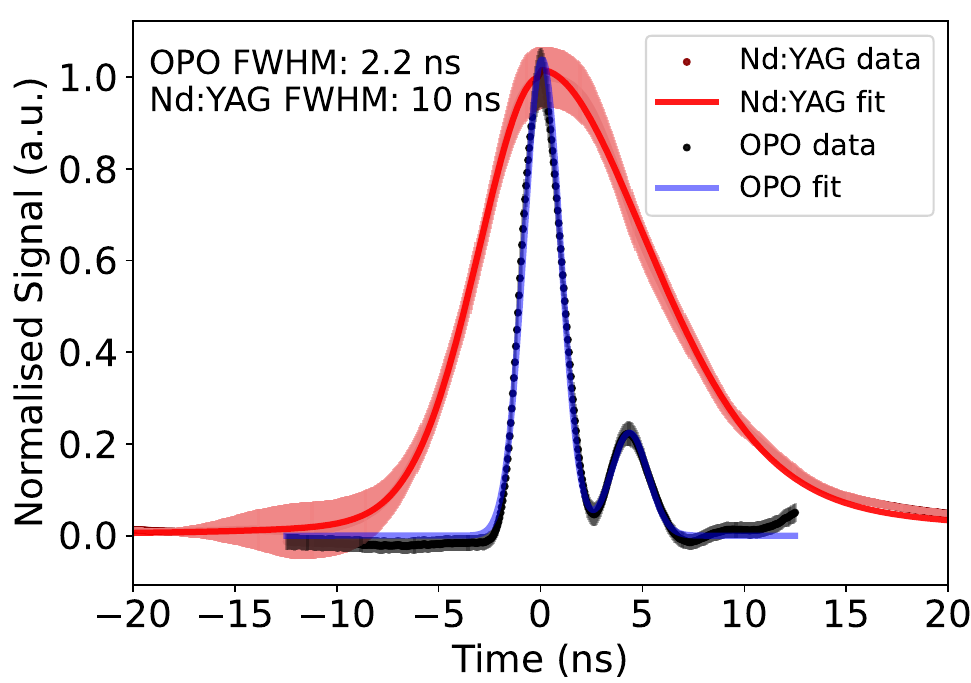}
        \caption{Averaged pulse shape over 1000 pulses with respective standard deviation on each point (surrounding band) for both lasers. OPO laser (black) and Nd:YAG laser (red). A reflection in the setup introduces a secondary peak in the OPO signal, which was not accounted for in the determination of the full width at half maximum (FWHM). }
    \end{subfigure}
    \caption{(a) Pulse timing profiles of the lasers and their (b) pulse-to-pulse timing stability (jitter)}
    \label{fig:laserpulse}
\end{figure*}

\smallskip
\item For each lifetime measurement, the background signal was characterised under several distinct experimental conditions: with the ion beam blocked, with all lasers off, with only the excitation laser on, with only the ionisation laser on, and with both lasers on but far detuned from any resonance. In each case, the background level was extracted by taking the average number of counts on the detector over the full-time window of the corresponding dataset. The total mean background was then reconstructed based on the contributions from these measurements.
The mean components associated with the ion bean blocked and laser-off conditions were treated as linearly additive, as they originate from independent sources: residual background radiation in the beamline,  dark counts on the ion detector, or noise on the electronic signal for the first one and collisional excitation processes for the second. In contrast, the laser-induced backgrounds (Nd:YAG, OPO, or both) were not assumed to be additive, due to potential non-linear dependencies in contributions to the effect. Since the lasers were detuned  from any other known resonances, these backgrounds  are associated with non-resonant excitation processes, such as multiphotons effects. Under this assumption, the highest background level observed among the laser-on configurations was taken as a conservative upper bound for the laser-induced background in each measurement. Since the spatial overlap and alignment were individually optimised for each level measurement to maximise the signal, the efficiency and magnitude of laser-induced background contributions varied across datasets. Therefore, these components were treated independently in the analysis and fitted separately for each measurement condition. The total mean background $\langle B_\mathrm{total} \rangle$ can thus be expressed as
\[
\langle B_{\rm total} \rangle 
\,=\, \langle B_{\rm beamgate} \rangle 
+ \langle B_{\rm laseroff} \rangle 
+ \max \Big( \langle B_{\rm Nd:YAG} \rangle , \langle B_{\rm OPO} \rangle \Big)   \;  .
\]

To account for uncertainty in the background measurement, it is treated as a constrained parameter in the exponential decay fit rather than fixing it \emph{a priori}. The background was allowed to vary within the range defined by its experimental uncertainty, using the measured value as a prior estimate. This approach ensures proper uncertainty propagation, reduces bias in the decay parameters, and accounts for potential variations between background and signal acquisition conditions.

\smallskip
\item The systematic uncertainty arising from cascade feeding was evaluated by propagating the lifetimes of the lower-lying states. Since the total signal
\begin{equation}
N_{\mathrm{tot}}(t) \,=\, N_a(t) + \sum_j N_j(t) + C   \; ,
\end{equation}
depends explicitly on the cascade lifetimes $\tau_j$, uncertainties in these previously determined quantities influence the extracted upper-level lifetime $\tau_a$. In particular, through

\begin{equation}
    N_j(t) \,=\, N_0 \, b_{aj} \,
    \frac{\tau_j}{\tau_a - \tau_j}
    \pqty{\mathrm{e}^{-t/\tau_a}-\mathrm{e}^{-t/\tau_j}}   \;  ,
\end{equation}
variations in $\tau_j$ modify the time dependence of the cascade contribution and therefore the fitted value of $\tau_a$. 

\medskip
The propagation was evaluated numerically using a finite-difference approach. For each cascade level $j$, the lifetime was varied independently by $\pm \sigma_{\tau_j}$, and the decay curve was refitted to extract the corresponding values $\tau_a^{(+)}$ and $\tau_a^{(-)}$. The sensitivity of the upper-level lifetime to the parameter $\tau_j$ was then estimated as
\begin{equation}
\Delta \tau_a^{(j)} \,=\,
\frac{\tau_a^{(+)} - \tau_a^{(-)}}{2}   \;  ,
\end{equation}
which corresponds to the first-order propagation term
\begin{equation}
\Delta \tau_a^{(j)} 
\,\approx\, 
\left|\frac{\partial \tau_a}{\partial \tau_j}\right|
\sigma_{\tau_j}   \;  .
\end{equation}
Assuming independent cascade parameters, the total systematic uncertainty from cascade feeding was obtained by adding the individual contributions in quadrature,
\begin{equation}
\sigma_{\tau_a,\mathrm{sys}}^2 \,=\,
\sum_j \left( \Delta \tau_a^{(j)} \right)^2    \;  .
\end{equation}
This systematic contribution was then combined in quadrature with the statistical uncertainty from the fit to obtain the total uncertainty on $\tau_a$.
\end{itemize}

\begin{table}[t!]
\centering
\caption{(Left) E1 Decay channel and branching ratios of different $n\pr\,^2\PR_{\nicefrac{1}{2}}$ and $n\pr\,^2\PR_{\nicefrac{3}{2}}$ states of $^{221}$Fr. (Right) Lifetimes of Fr levels.}
\begin{tabular}{c c c cc c c||ccc}
\hline\hline
State & Decay Channel &   $b_{aj}$   && State & Decay Channel &  $b_{aj}$   & Level & $\tau$(ns) & source\\
\hline
$8\pr\,^2\PR_{3/2}$ & $\rightarrow$ $7\sr\,^2\SR_{\nicefrac{1}{2}}$ & 0.47 & & & & &$10\pr\,^2\PR_{3/2}$ & 362(7)[3] & this work\\
& $\rightarrow$ $6\dr\,^2\DR_{\nicefrac{3}{2}}$ & 0.01 && & &&$10\pr\,^2\PR_{1/2}$ & 553(21)[8]   & this work\\
& $\rightarrow$ $6\dr\,^2\DR_{\nicefrac{5}{2}}$ & 0.11 && & &&$10\sr\,^2\SR_{1/2}$ & 230.4(46) & \cite{Weijngaarden_lifetimes_1999} \\
& $\rightarrow$ $8\sr\,^2\SR_{\nicefrac{1}{2}}$ & 0.42 && & &&$9\pr\,^2\PR_{3/2}$  & 179(5)[3] & this work \\[1ex]
\cline{1-7}
$9\pr\,^2\PR_{\nicefrac{1}{2}}$ & $\rightarrow$ $7\sr\,^2\SR_{\nicefrac{1}{2}}$ & 0.07 && $9\pr\,^2\PR_{\nicefrac{3}{2}}$ & $\rightarrow$ $7\sr\,^2\SR_{\nicefrac{1}{2}}$ & 0.36 &$9\pr\,^2\PR_{1/2}$  & 329(6)[2] & this work\\
& $\rightarrow$ $6\dr\,^2\DR_{\nicefrac{3}{2}}$ & 0.24 && & $\rightarrow$ $6\dr\,^2\DR_{3/2}$ & 0.01 & $9\sr\,^2\SR_{1/2}$  & 107.53(90) & \cite{aubin_lifetimes_2004} \\
& $\rightarrow$ $8\sr\,^2\SR_{\nicefrac{1}{2}}$ & 0.12 && & $\rightarrow$ $6\dr\,^2\DR_{\nicefrac{5}{2}}$ & 0.08 & $8\pr\,^2\PR_{3/2}$  & 83.5(15)  & \cite{aubin_lifetimes_2004}\\
& $\rightarrow$ $7\dr\,^2\DR_{\nicefrac{3}{2}}$ & 0.29 && & $\rightarrow$ $8\sr\,^2\SR_{\nicefrac{1}{2}}$ & 0.21 & $8\pr\,^2\PR_{1/2}$  & 149.3(35) & \cite{aubin_lifetimes_2004}\\
& $\rightarrow$ $9\sr\,^2\SR_{\nicefrac{1}{2}}$ & 0.28 && & $\rightarrow$ $7\dr\,^2\DR_{\nicefrac{3}{2}}$ & 0.01&$8\sr\,^2\SR_{1/2}$  & 53.30(44) & \cite{gomez_lifetime_2005} \\ 
& & && & $\rightarrow$ $7\dr\,^2\DR_{\nicefrac{5}{2}}$ & 0.11&$8\dr\,^2\DR_{5/2}$  & 95.9(19)  & \cite{Weijngaarden_lifetimes_1999} \\
& & && & $\rightarrow$ $9\sr\,^2\SR_{\nicefrac{1}{2}}$ & 0.23& $8\dr\,^2\DR_{3/2}$  & 99.9(20)  & \cite{Weijngaarden_lifetimes_1999}\\[1ex]
\cline{1-7}
$10\pr\,^2\PR_{\nicefrac{1}{2}}$ & $\rightarrow$ $7\sr\,^2\SR_{\nicefrac{1}{2}}$ & 0.04 &&  $10\pr\,^2\PR_{3/2}$ & $\rightarrow$ $7\sr\,^2\SR_{\nicefrac{1}{2}}$ & 0.30&$7\pr\,^2\PR_{3/2}$  & 21.02(11) & \cite{simsarian_lifetime_1998} \\
& $\rightarrow$ $6\dr\,^2\DR_{\nicefrac{3}{2}}$ & 0.23 && & $\rightarrow$ $6\dr\,^2\DR_{\nicefrac{3}{2}}$ & 0.01&$7\pr\,^2\PR_{1/2}$  & 29.45(11) & \cite{simsarian_lifetime_1998}\\
& $\rightarrow$ $8\sr\,^2\SR_{\nicefrac{1}{2}}$ & 0.09 && & $\rightarrow$ $6\dr\,^2\DR_{\nicefrac{5}{2}}$ & 0.07&$7\dr\,^2\DR_{5/2}$  & 67.7(29)  & \cite{grossman_lifetime_2000}\\
& $\rightarrow$ $7\dr\,^2\DR_{\nicefrac{3}{2}}$ & 0.20 && & $\rightarrow$ $8\sr\,^2\SR_{\nicefrac{1}{2}}$ & 0.18&$7\dr\,^2\DR_{3/2}$  & 73.6(3)   & \cite{grossman_lifetime_2000}\\
& $\rightarrow$ $9\sr\,^2\SR_{\nicefrac{1}{2}}$ & 0.11 && & $\rightarrow$ $7\dr\,^2\DR_{\nicefrac{3}{2}}$ & 0.01&$6\dr\,^2\DR_{5/2}$  & 1915(38)  & \cite{Weijngaarden_lifetimes_1999}\\ 
& $\rightarrow$ $8\dr\,^2\DR_{\nicefrac{3}{2}}$ & 0.15 && & $\rightarrow$ $7\dr\,^2\DR_{\nicefrac{5}{2}}$ & 0.07&$6\dr\,^2\DR_{3/2}$  & 559.2(11) & \cite{Weijngaarden_lifetimes_1999}\\
& $\rightarrow$ $10\sr\,^2\SR_{\nicefrac{1}{2}}$& 0.18 && & $\rightarrow$ $9\sr\,^2\SR_{\nicefrac{1}{2}}$ & 0.15& $7\sr\,^2\SR_{1/2}$  & --- &  \\
& & && & $\rightarrow$ $8\dr\,^2\DR_{\nicefrac{3}{2}}$ & 0.01\\
& & && & $\rightarrow$ $8\dr\,^2\DR_{\nicefrac{5}{2}}$ & 0.06\\
& & && & $\rightarrow$ $10\sr\,^2\SR_{\nicefrac{1}{2}}$ & 0.16\\[1ex]
\hline\hline
\end{tabular}
\label{Tab-BR}
\end{table}

\subsection{Lifetimes, Branching ratios and dipole matrix elements}

Calculations of the electric dipole (E1) matrix elements were performed using the RCCSDT method (see section 3.2 above), and the corresponding transition rates and lifetimes were determined from these matrix elements. The calculated lifetimes were then compared with the experimental values to assess the accuracy of the E1 matrix elements, as shown in Tab.\ \ref{tab:tablefreq}. 
The comparison between experimental and theoretical lifetimes shows agreement within uncertainties for all states: the 9p and 10p levels exhibit relative deviations of 4\% or less, with the largest deviation observed for the $9\pr\,{}^2\PR_{\nicefrac{3}{2}}$ state and the smallest for $10\pr\,{}^2\PR_{\nicefrac{3}{2}}$. Disrepancies may result from incomplete treatment of electron correlation and the extended spatial distribution of high-lying Rydberg orbitals. The RCC calculations of alkali-metal atoms have shown that high-lying states are sensitive to the spatial extent of the wavefunctions and electron correlation, which the coupled-cluster singles-and-doubles (CCSD) approximation may not adequately capture~\cite{Pal__Rel_2007,safronova_crit_2011}.
The agreement between the experimental and theoretical values indicates that the RCCSDT method yields accurate E1 matrix elements.

The transition probability (in s$^{-1}$) of an atomic state ($|\Psi_\ir \rangle$) to a lower state ($|\Psi_\fr \rangle$) due to the E1 decay channel is given by
\begin{eqnarray}\label{decay}
    A_{\ir \rightarrow \fr} \,=\, \frac{2.02613 \times 10^{15}}{(2J_\ir +1) \,\lambda_{\ir \rightarrow \fr}^3} \: |\langle J_\ir || D || J_\fr \rangle|^2   \;  ,
\end{eqnarray}
where $\lambda_{\ir \rightarrow \fr}$ is the transition wavelength (in nm) and $\langle J_i || D || J_f \rangle$ is the reduced E1 transition matrix element (in atomic units [a.u.]) with the angular momentum of the states as $J$. The lifetime (in s) of the state $|\Psi_\ir \rangle$ is then determined by 
\begin{eqnarray}
\tau_\ir \,=\, \frac{1} {\sum\limits_\fr A_{\ir \rightarrow \fr} }    \;  .
\end{eqnarray}

The branching ratios ($b_{aj}$) for each state are listed in Tab.~\ref{Tab-BR}. According to Eq.\ (\ref{decay}), the transition probability depends on the cube of the energy difference associated with the decay channel. As the principal quantum number $n$ increases, the energy spacing between adjacent states becomes smaller. For any $n\pr\,^2\PR_{\nicefrac{1}{2}}$ or $n\pr\,^2\PR_{\nicefrac{3}{2}}$ state, the primary decay path is to the $n\sr\,^2\SR_{\nicefrac{1}{2}}$ state. However, at higher $n$, the energy gap between the $n\pr$ and $n\sr$ levels decreases, leading to a reduced transition rate $A$, and thus longer lifetimes. This trend is also evident from the ($b_{aj}$) in Tab.~\ref{Tab-BR}. Additionally, the E1 matrix element between high $n\pr$ states and lower $n\sr$ states is relatively small; although the energy gap in these cases is larger, it does not significantly affect the lifetime of the corresponding $\PR_{\nicefrac{1}{2}}$ and $\PR_{\nicefrac{3}{2}}$ states. However, for the $n\pr\,^2\PR_{\nicefrac{3}{2}}$ state, the E1 transition matrix element overlap with the $7\sr\,^2\SR_{{\nicefrac{1}{2}}}$ state is significantly larger, which leads to shorter lifetimes compared to its $\PR_{\nicefrac{1}{2}}$ counterparts. See also in Tab.\ \ref{Tab-BR} the values of lifetimes used in the different cascade fit with their uncertainties.

\subsection{Discussion on the experimental and theoretical data}
The comparison between the theoretical and experimental radiative lifetimes shows good overall agreement for the states in the 9p and 10p configurations. The differences between RCCSDT and experimental values remain within the combined uncertainties for all four levels, with relative discrepancies of approximately \(3.5\%\), \(3.9\%\), \(3.8\%\), and \(1.7\%\) for $9\pr\,{}^2\PR_{\nicefrac{1}{2}}$, $9\pr\,{}^2\PR_{\nicefrac{3}{2}}$, $10\pr\,{}^2\PR_{\nicefrac{1}{2}}$, and $10\pr\,{}^2\PR_{\nicefrac{3}{2}}$, respectively. The largest experimental uncertainty is observed for the $10\pr\,{}^2\PR_{\nicefrac{1}{2}}$ lifetime, where the statistical and systematic contributions lead to an uncertainty of about \(4\%\). 
Since the radiative lifetime depends on the inverse of the total transition rate, even small uncertainties in the E1 transition matrix elements can lead to noticeable changes in the calculated lifetimes. On the theoretical side, residual higher-order electron-correlation effects beyond triple excitations, incomplete treatment of core-valence correlations, and finite-basis limitations may contribute to the observed differences. These effects can be particularly relevant for francium, where relativistic and correlation effects strongly influence the electronic wavefunctions. Smaller contributions from QED corrections may also become relevant at the present level of accuracy. In contrast to the excitation energies, for which the theoretical values show a more systematic offset with respect to experiment, the lifetime discrepancies exhibit no clear systematic trend: the theoretical values are higher than the experimental ones for the $9\pr\,{}^2\PR_{\nicefrac{1}{2}}$ and $10\pr\,{}^2\PR_{\nicefrac{3}{2}}$ states, but lower for the $9\pr\,{}^2\PR_{\nicefrac{3}{2}}$ and $10\pr\,{}^2\PR_{\nicefrac{1}{2}}$ states. This alternating behaviour suggests that the remaining differences are more likely associated with state-dependent uncertainties in the calculated transition rates and experimental measurements, rather than with a single systematic shift. Overall, the theoretical and experimental lifetimes remain in good agreement, within the combined uncertainties.

\begin{figure}[ht!]
    \centering

    \begin{minipage}[c]{0.45\textwidth}
        \centering
        \includegraphics[width=\textwidth]{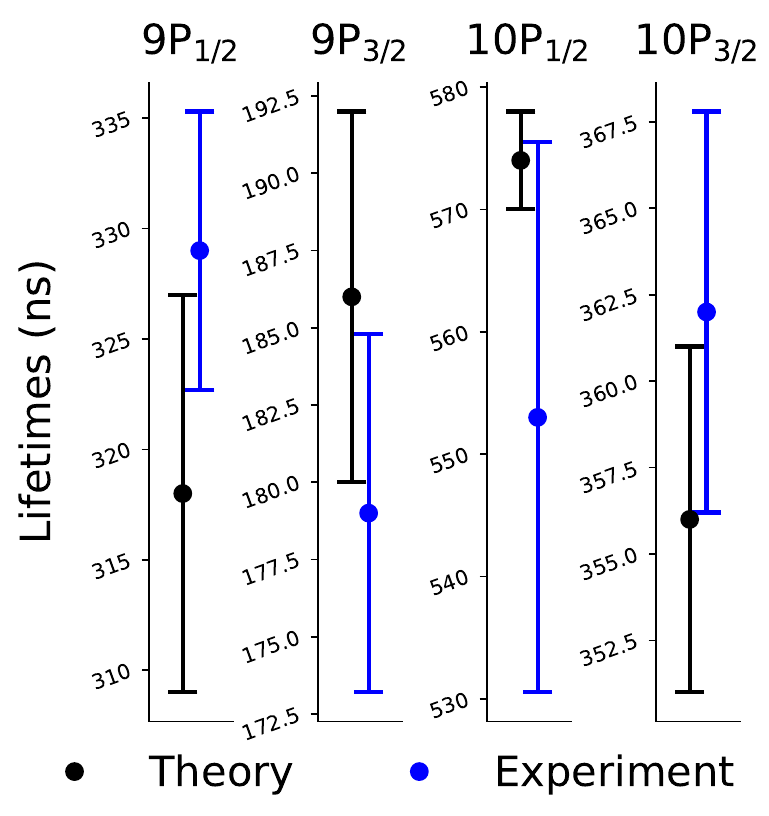}
        \captionof{figure}{Comparison of the experimental and RCCSDT radiative lifetimes. Vertical error bars represent the corresponding theoretical and experimental uncertainties.}
        \label{fig:lifetimescomp}
    \end{minipage}
    \hspace{0.02\textwidth}
    \begin{minipage}[c]{0.50\textwidth}
        \centering
        \captionof{table}{Measured and calculated radiative lifetimes for $^{221}$Fr.}
        \label{tab:tablelifetimes}

        \begin{tabular}{@{}lcc@{}}
            \toprule
            Level & \multicolumn{2}{c}{Lifetime (ns)} \\
            \cmidrule{2-3}
            & Theory & Experiment \\
            \midrule

            $8\pr\,^2\PR_{\nicefrac{3}{2}}$
            & $80.5(2.9)^\mathrm{a}$
            & $81.1(1.7)^\mathrm{a}$ \\
            &
            $81.4^\mathrm{b}$
            & $83.5(1.5)^\mathrm{e}$ \\
            &
            $81.8(2.3)^\mathrm{c}$
            & \\
            &
            $61.13^\mathrm{d}$
            & \\[1mm]

            $9\pr\,^2\PR_{\nicefrac{1}{2}}$
            & $318(9)^\mathrm{a}$
            & $329(6)[2]^\mathrm{a}$ \\
            &
            $319(10)^\mathrm{c}$
            & \\
            &
            $233.4^\mathrm{d}$
            & \\[1mm]

            $9\pr\,^2\PR_{\nicefrac{3}{2}}$
            & $186(6)^\mathrm{a}$
            & $179(5)[3]^\mathrm{a}$ \\
            &
            $186(2)^\mathrm{c}$
            & \\
            &
            $130.7^\mathrm{d}$
            & \\[1mm]

            $10\pr\,^2\PR_{\nicefrac{1}{2}}$
            & $574(4)^\mathrm{a}$
            & $553(21)[8]^\mathrm{a}$ \\
            &
            $566(7)^\mathrm{c}$
            & \\
            &
            $424.3^\mathrm{d}$
            & \\[1mm]

            $10\pr\,^2\PR_{\nicefrac{3}{2}}$
            & $356(5)^\mathrm{a}$
            & $362(5)[3]^\mathrm{a}$ \\
            &
            $346(5)^\mathrm{c}$
            & \\
            &
            $231.3^\mathrm{d}$
            & \\

            \bottomrule
        \end{tabular}

        \vspace{2mm}

        \footnotesize
        Refs. $^\mathrm{a}$This work;
        $^\mathrm{b}$Estimated using the E1 matrix elements from \cite{Roberts2023};
        $^\mathrm{c}$\cite{Tang2017};
        $^\mathrm{d}$\cite{Weijngaarden_lifetimes_1999};
        $^\mathrm{e}$\cite{aubin_lifetimes_2004}.
        ( ) statistical; [ ] systematic uncertainties, given when significant.
        See Tab.~\ref{tab:lifetimeuncertainties}.
    \end{minipage}

\end{figure}

\section{Conclusion}
The absolute wavenumbers, fine-structure splittings, and radiative lifetimes of the francium 9p and 10p states were measured using collinear resonance ionisation spectroscopy on a $^{221}$Fr beam, with centroids extracted from fits of resonance spectra and lifetimes determined from time-resolved decay curves. The analyses yield a consistent set of experimental parameters for these excited states, with uncertainties quantified from statistical fluctuations, laser systematics, and cascade contributions in the decay model. New calculations using the singles, doubles and triples approximated relativistic coupled-cluster (RCCSDT) method were performed in parallel to provide \emph{ab initio} predictions of excitation energies and radiative lifetimes, enabling direct comparison with the measured values.

Two complementary conclusions can be drawn. First, while the RCCSDT calculations exhibit a systematic offset in the absolute binding energies, the relative excitation energies are reproduced with high accuracy. This indicates that the dominant missing contributions, such as higher-order excitations beyond triples, contributions from high-angular-momentum orbitals, and QED effects enter primarily as an approximately state-independent shift to the valence removal energy. 
The accurate reproduction of relative level spacings therefore demonstrates that the RCCSDT method captures the essential state-dependent correlation physics in francium.

Second, the agreement between calculated and measured lifetimes of all measured states provides an independent and more stringent validation of the theoretical wave functions. Since excited-state lifetimes are directly determined by electric-dipole (E1) transition matrix elements, this agreement indicates that the RCCSDT approach yields accurate E1 matrix elements, which can serve as reliable theoretical inputs for atomic parity violation (PV) calculations in francium. A direct connection between the accuracy of the E1 matrix elements and PV calculations can be illustrated by the ($7\sr\rightarrow8\sr$) transition. The PV electric-dipole amplitude for this transition is evaluated as
\begin{equation}
  E1_{\mathrm{PV}} \,=\,
\sum_n \left[
\frac{\langle 7\sr\,{}^2\SR_{\nicefrac{1}{2}}|H_{\mathrm{PV}}|n\pr\,{}^2\PR_{\nicefrac{1}{2}}\rangle
\langle n\pr\,{}^2\PR_{\nicefrac{1}{2}}|D|8\sr\,{}^2\SR_{\nicefrac{1}{2}}\rangle}
{E_{7\sr\,{}^2\SR_{\nicefrac{1}{2}}}-E_{n\pr\,{}^2\PR_{\nicefrac{1}{2}}}}
+
\frac{\langle 7\sr\,{}^2\SR_{\nicefrac{1}{2}}|D|n\pr\,{}^2\PR_{\nicefrac{1}{2}}\rangle
\langle n\pr\,{}^2\PR_{\nicefrac{1}{2}}|H_{\mathrm{PV}}|8\sr\,{}^2\SR_{\nicefrac{1}{2}}\rangle}
{E_{8\sr\,{}^2\SR_{\nicefrac{1}{2}}}-E_{n\pr\,{}^2\PR_{\nicefrac{1}{2}}}}
\right]   \; ,  
\end{equation}
where the summation extends over the intermediate ($n\pr\,{}^2\PR_{\nicefrac{1}{2}}$) states ($n=7,8,9, \, \ldots$). The PV amplitude contains the E1 matrix elements ($\langle n\pr\,{}^2\PR_{\nicefrac{1}{2}}|D|8\sr\,{}^2\SR_{\nicefrac{1}{2}}\rangle$) and ($\langle 7\sr\,{}^2\SR_{\nicefrac{1}{2}}|D|n\pr\,{}^2\PR_{\nicefrac{1}{2}}\rangle$), in addition to the weak-interaction matrix elements. The accuracy of these E1 matrix elements is therefore directly relevant to the reliability of the calculated PV amplitude. Experimental E1 matrix elements, which can be determined from atomic lifetime measurements or electric-dipole polarisability measurements, provide an essential benchmark for testing the accuracy of the many-body calculations. Thus, the agreement between the E1 matrix elements obtained from our lifetime measurements and the RCCSDT calculations provides direct experimental validation of an important set of theoretical inputs required for PV studies in francium
\cite{aoki2018parity,kastberg2019}.\\ 
Together, these results confirm that calculations using the RCCSDT method provides a reliable and internally-consistent description of both the atomic energy structure and transition properties of francium, supporting its use in precision spectroscopy and atomic PV studies. Future work will include results from ongoing measurements of the higher members of the p series as well as the 6d and higher angular momentum states, while a more precise determination of isotope shifts remains an important goal. As stated earlier, spectroscopic data of any kind is very sparse for Fr states below the ones lowest in energy. Thus, such measurements will not only complete the spectroscopic picture of francium but also provide essential input data for tests of fundamental symmetries.\\

\textbf{Data Availability Statement}: The data supporting the findings of this study are openly available at \cite{lassegues_dataset_2026}.

\newpage

\bibliographystyle{spphys}
\bibliography{ref_opt}

@misc{lassegues_dataset_2026,
	title = {Dataset for "{Energies} and lifetimes of the 9p and 10p fine-structure doublets of atomic francium"},
	doi = {10.5281/zenodo.19555372},
	urldate = {2026-04-14},
	publisher = {Zenodo},
	author = {Lassègues, P. and others},
	month = apr,
	year = {2026},
}

@article{CMS2021Angular,
  author  = {{CMS Collaboration}},
  title   = {Angular analysis of the decay $B^{+} \to K^{*}(892)^{+}\mu^{+}\mu^{-}$ in proton-proton collisions at $\sqrt{s}=8$ TeV},
  journal = {J. High Energy Phys.},
  volume  = {2021},
  number  = {4},
  pages   = {124},
  year    = {2021},
  doi     = {10.1007/JHEP04(2021)124},
}

@article{lalanne_2026,
  title = {Resonance laser ionization spectroscopy of stable chromium isotopes},
  author = {Lalanne, L. and Mancheva, R. and Athanasakis-Kaklamanakis, M. and Heines, M. and Koszorus, {\'A}. and Liu, Y. C. and Reilly, J. and Bernerd, C. and van den Borne, B. and Chrysalidis and others},
  journal = {Phys. Rev. A},
  volume = {114},
  issue = {3},
  pages = {032811},
  numpages = {11},
  year = {2026},
  month = {Sep},
  publisher = {American Physical Society},
  doi = {https://link.aps.org/doi/10.1103/4kpl-b7cs}
}

@article{
Roussy2023,
author = {Tanya S. Roussy  and Luke Caldwell  and Trevor Wright  and William B. Cairncross  and Yuval Shagam  and Kia Boon Ng  and Noah Schlossberger  and Sun Yool Park  and Anzhou Wang  and others},
title = {An improved bound on the electron’s electric dipole moment},
journal = {Science},
volume = {381},
number = {6653},
pages = {46-50},
year = {2023},
doi = {https://doi.org/10.1126/science.adg4084},

}

@article{ATHANASAKISKAKLAMANAKIS202386,
title = {Voltage scanning and technical upgrades at the Collinear Resonance Ionization Spectroscopy experiment},
journal = {Nuclear Instruments and Methods in Physics Research Section B: Beam Interactions with Materials and Atoms},
volume = {541},
pages = {86-89},
year = {2023},
issn = {0168-583X},
doi = {https://doi.org/10.1016/j.nimb.2023.04.054},
author = {Michail Athanasakis-Kaklamanakis and Jordan R. Reilly and Ágota Koszorus and Shane G. Wilkins and Louis Lalanne and Sarina Geldhof and Miranda Nichols and Quanjun Wang and Bram {van den Borne} and David Chorlton and others},

}

@article{chupp2019,
  title = {Electric dipole moments of atoms, molecules, nuclei, and particles},
  author = {Chupp, T. E. and Fierlinger, P. and Ramsey-Musolf, M. J. and Singh, J. T.},
  journal = {Rev. Mod. Phys.},
  volume = {91},
  issue = {1},
  pages = {015001},
  numpages = {55},
  year = {2019},
  month = {Jan},
  publisher = {American Physical Society},
  doi = {https://link.aps.org/doi/10.1103/RevModPhys.91.015001}
}

@article{Arrowsmith-Kron_2024,
doi = {https://doi.org/10.1088/1361-6633/ad1e39},
year = {2024},
month = {jul},
publisher = {IOP Publishing},
volume = {87},
number = {8},
pages = {084301},
author = {Arrowsmith-Kron, Gordon and Athanasakis-Kaklamanakis, Michail and Au, Mia and Ballof, Jochen and Berger, Robert and Borschevsky, Anastasia and Breier, Alexander A and Buchinger, Fritz and Budker, Dmitry and Caldwell, Luke and others},
title = {Opportunities for fundamental physics research with radioactive molecules},
journal = {Reports on Progress in Physics},
}

@misc{nudat3,
  author       = {{National Nuclear Data Center}},
  title        = {{NuDat} database},
  howpublished = {\url{https://bnl.gov}},
  year         = {2026}
}

@article{PhysRevA.108.062815,
  title = {Candidate molecules for next-generation searches of hadronic charge-parity violation},
  author = {Marc, Aur\'elien and Hubert, Micka\"el and Fleig, Timo},
  journal = {Phys. Rev. A},
  volume = {108},
  issue = {6},
  pages = {062815},
  numpages = {7},
  year = {2023},
  month = {Dec},
  publisher = {American Physical Society},
  doi = {10.1103/PhysRevA.108.062815},
  url = {https://link.aps.org/doi/10.1103/PhysRevA.108.062815}
}

@article{Gwinner_2022,
doi = {10.1088/2058-9565/ac4424},
url = {https://doi.org/10.1088/2058-9565/ac4424},
year = {2022},
month = {jan},
publisher = {IOP Publishing},
volume = {7},
number = {2},
pages = {024001},
author = {Gwinner, G and Orozco, L A},
title = {Studies of the weak interaction in atomic systems: towards measurements of atomic parity non-conservation in francium},
journal = {Quantum Science and Technology}
}

@article{PhysRevA.63.062101,
  title = {Calculations of parity-nonconserving $s\ensuremath{-}d$ amplitudes in Cs, Fr, ${\mathrm{Ba}}^{+},$ and ${\mathrm{Ra}}^{+}$},
  author = {Dzuba, V. A. and Flambaum, V. V. and Ginges, J. S. M.},
  journal = {Phys. Rev. A},
  volume = {63},
  issue = {6},
  pages = {062101},
  numpages = {8},
  year = {2001},
  month = {May},
  publisher = {American Physical Society},
  doi = {https://doi.org/10.1103/PhysRevA.63.062101}
}

@article{vidal2019standardmodelphysicshllhc,
    author = {X. Cid Vidal and M. D'Onofrio and P. J. Fox and R. Torre and K. A. Ulmer and A. Aboubrahim and A. Albert and J. Alimena and B. C. Allanach and others},
    doi = {https://doi.org/10.48550/arXiv.1812.07831},
    primaryclass = {hep-ph},
    journal = {CERN-LPCC},
    title = {Beyond the Standard Model Physics at the HL-LHC and HE-LHC},
    year = {2019}
}

@article{Tarbutt_2009,
    author = {Tarbutt, M.R. and Hudson, J.J. and Sauer, B.E. and Hinds, E.A.},
    doi = {https://doi.org/10.1039/b820625b},
    issn = {1364-5498},
    journal = {Faraday Discussions},
    pages = {37},
    publisher = {Royal Society of Chemistry (RSC)},
    title = {Prospects for measuring the electric dipole moment of the electron using electrically trapped polar molecules},
    volume = {142},
    year = {2009}
}

@article{sakemi_search_2011,
    author = {Sakemi, Y. and Harada, K. and Hayamizu, T. and Itoh, M. and Kawamura, H. and Liu, S. and Nataraj, H. S. and Oikawa, A. and others},
    date = {2011-07},
    doi = {https://doi.org/10.1088/1742-6596/302/1/012051},
    journal = {J. Phys.: Conf. Ser.},
    number = {1},
    title = {Search for a permanent {EDM} using laser cooled radioactive atom},
    volume = {302},
    year = {2011}
}

@article{aubin_atomic_2013,
    author = {Aubin, S. and Behr, J.A. and Collister, R. and Flambaum, V.V. and Gomez, E. and Gwinner, G. and Jackson, K.P. and others},
    doi = {https://doi.org/10.1007/s10751-013-0797-6},
    journal = {Hyperfine Interact},
    pages = {163--171},
    title = {Atomic parity non-conservation: the francium anapole project of the {FrPNC} collaboration at {TRIUMF}},
    volume = {214},
    year = {2013}
}

@article{Gomez_2007,
    author = {Gomez, E. and Aubin, S. and Sprouse, G. D. and Orozco, L. A. and DeMille, D. P.},
    doi = {https://doi.org/10.1103/PhysRevA.75.033418},
    issue = {3},
    journal = {Phys. Rev. A},
    month = {Mar},
    numpages = {10},
    pages = {033418},
    publisher = {American Physical Society},
    title = {Measurement method for the nuclear anapole moment of laser-trapped alkali-metal atoms},
    volume = {75},
    year = {2007}
}

@article{aoki2018parity,
    author = {Aoki, T. and Torii, Y. and Sahoo, B.K. and Das, B.P. and Harada, K. and Hayamizu, T. and Sakamoto, K. and Kawamura, H. and Inoue, T. and Uchiyama, A. and others},
    doi = {https://doi.org/10.1007/s00340-017-6673-3},
    journal = {Appl. Phys. B},
    pages = {509--527},
    publisher = {Springer},
    title = {Parity-Nonconserving Interaction-Induced Light Shifts in the--Transition of the Ultracold Atoms to Probe New Physics Beyond the Standard Model},
    year = {2018}
}

@article{kastberg2019,
    author = {Kastberg, A. and Aoki, T. and Sahoo, B.K. and Sakemi, Y. and Das, B.P.},
    doi = {https://doi.org/10.1103/PhysRevA.100.050101},
    journal = {Physical Review A},
    number = {5},
    pages = {050101},
    publisher = {APS},
    title = {Optical-lattice-based method for precise measurements of atomic parity violation},
    volume = {100},
    year = {2019}
}

@article{Zhao_1997,
    author = {Zhao, W.Z. and Simsarian, J.E. and Orozco, L.A. and Shi, W. and Sprouse, G.D.},
    doi = {https://doi.org/10.1103/PhysRevLett.78.4169},
    issue = {22},
    journal = {Phys. Rev. Lett.},
    month = {Jun},
    numpages = {0},
    pages = {4169--4172},
    publisher = {American Physical Society},
    title = {Measurement of the $7\mathit{p}{}^{2}{\mathit{P}}_{3/2}$ Level Lifetime in Atomic Francium},
    volume = {78},
    year = {1997}
}

@article{safronova_2007,
    author = {Safronova, U.I. and Johnson, W.R. and Safronova, M.S.},
    doi = {https://doi.org/10.1103/PhysRevA.76.042504},
    issue = {4},
    journal = {Phys. Rev. A},
    month = {Oct},
    numpages = {15},
    pages = {042504},
    publisher = {American Physical Society},
    title = {Excitation energies, polarizabilities, multipole transition rates, and lifetimes of ions along the francium isoelectronic sequence},
    volume = {76},
    year = {2007}
}

@article{sahoo_correlation_2015,
    author = {Sahoo, B.K. and Nandy, D.K. and Das, B.P. and Sakemi, Y.},
    date = {2015-04},
    doi = {https://doi.org/10.1103/PhysRevA.91.042507},
    journal = {Phys. Rev. A},
    journaltitle = {Phys. Rev. A},
    number = {4},
    pages = {042507},
    title = {Correlation trends in the hyperfine structures of 210,212 Fr},
    volume = {91},
    year = {2015}
}

@article{Vajed_1982,
    author = {Vajed-Samii, M. and Andriessen, J. and Das, B.P. and Ray, S.N. and Lee, T. and Das, T.P.},
    doi = {https://doi.org/10.1103/PhysRevLett.49.1800},
    issue = {24},
    journal = {Phys. Rev. Lett.},
    month = {Dec},
    numpages = {0},
    pages = {1800--1800},
    publisher = {American Physical Society},
    title = {Accurate Values of Nuclear Magnetic Moments of Francium Isotopes.},
    volume = {49},
    year = {1982}
}

@article{Owusu_1997,
    author = {Owusu, A. and Dougherty, R.W. and Gowri, G. and Das, T.P. and Andriessen, J.},
    doi = {https://doi.org/10.1103/PhysRevA.56.305},
    issue = {1},
    journal = {Phys. Rev. A},
    month = {Jul},
    numpages = {0},
    pages = {305--309},
    publisher = {American Physical Society},
    title = {Relativistic many-body investigation of hyperfine interactions in excited $S$ states of alkali metals: Francium and potassium},
    volume = {56},
    year = {1997}
}

@misc{NIST-ASD,
    author = {Kramida, A.\ and
Ralchenko, Yu.\ and 
Reader, J.\ and 
{NIST ASD Team}},
    title = {{NIST} {Atomic Spectra Database}},
    url = {https://physics.nist.gov/asd},
    year = {2024}
}

@article{biemont_theoretical_1998,
    author = {Biémont, E. and Quinet, P. and Van Renterghem, V.},
    doi = {https://doi.org/10.1088/0953-4075/31/24/012},
    journal = {J. Phys. B: At. Mol. Opt. Phys.},
    number = {24},
    pages = {5301},
    title = {Theoretical investigation of neutral francium},
    volume = {31},
    year = {1998}
}

@article{das_study_2019,
    author = {Das, M. and Pradhan, A. C.},
    date = {2019-07},
    doi = {https://doi.org/10.1088/1361-6455/aafdd2},
    journal = {J. Phys. B: At. Mol. Opt. Phys.},
    number = {15},
    pages = {155001},
    title = {Study of atomic spectroscopy and hyperfine structure of francium (Fr) isotopes using relativistic Fock space multireference coupled-cluster method},
    volume = {52},
    year = {2019}
}

@article{Weijngaarden_lifetimes_1999,
    author = {W.A. {Van Wijngaarden} and 
J. Xia},
    doi = {https://doi.org/10.1016/S0022-4073(98)00028-4},
    journal = {J. Quant. Spec. Rad. Tr.},
    number = {4},
    pages = {557--561},
    title = {Lifetimes and polarizabilities of low lying S, P and D states of francium},
    volume = {61},
    year = {1999}
}

@article{kugler_isolde_1993,
    author = {Kugler, E.},
    doi = {https://doi.org/10.1016/0168-583X(93)95355-9},
    journal = {Nucl. Instrum. Methods Phys. Res. B},
    number = {1},
    pages = {322--325},
    title = {The ISOLDE facility at the CERN PS Booster},
    volume = {79},
    year = {1993}
}

@article{Catherall_2017,
    author = {Catherall, R. and Andreazza, W. and Breitenfeldt, M. and Dorsival, A. and Focker, G.J. and Gharsa, T.P. and Giles, T.J. and Grenard, J-L. and others},
    doi = {https://doi.org/10.1088/1361-6471/aa7eba},
    journal = {J. Phys. G: Nucl. Part. Phys.},
    month = {aug},
    number = {9},
    pages = {094002},
    publisher = {IOP Publishing},
    title = {The ISOLDE facility},
    volume = {44},
    year = {2017}
}

@article{giles_high_2003,
    author = {Giles, T.J. and Catherall, R. and Fedosseev, V. and Georg, U. and Kugler, E. and Lettry, J. and Lindroos, M.},
    date = {2003-05-01},
    doi = {https://doi.org/10.1016/S0168-583X(02)02119-5},
    issn = {0168-583X},
    journal = {Nucl. Instrum. Methods Phys. Res. Sect. B},
    pages = {497--501},
    title = {The high resolution spectrometer at {ISOLDE}},
    urldate = {2024-06-19},
    volume = {204},
    year = {2003}
}

@article{mane_ion_2009,
    author = {Mané, E. and Billowes, J. and Blaum, K. and Campbell, P. and Cheal, B. and Delahaye, P. and Flanagan, K. T. and Forest, D. H. and Franberg, H. and Geppert, C. and others},
    doi = {https://doi.org/10.1140/epja/i2009-10828-0},
    issn = {1434-6001, 1434-601X},
    journal = {Eur. Phys. J. A},
    journaltitle = {The European Physical Journal A},
    langid = {english},
    number = {3},
    pages = {503--507},
    title = {An ion cooler-buncher for high-sensitivity collinear laser spectroscopy at {ISOLDE}},
    urldate = {2024-06-19},
    volume = {42},
    year = {2009}
}

@article{Koszorus2020,
    author = {A. Koszorus and J. Billowes and C.L. Binnersley and M.L. Bissell and T.E. Cocolios and B. S. Cooper and R.P. de Groote and G.J. Farooq-Smith and V.N. Fedosseev and K.T. Flanagan and others},
    doi = {https://doi.org/10.1016/j.nimb.2019.04.043},
    issn = {0168-583X},
    journal = {Nucl. Instrum. Methods Phys. Res. B},
    pages = {398-402},
    title = {Resonance ionization schemes for high resolution and high efficiency studies of exotic nuclei at the CRIS experiment},
    volume = {463},
    year = {2020}
}

@article{Vernon2020,
    author = {A.R. Vernon and R.P. {de Groote} and J. Billowes and C.L. Binnersley and T.E. Cocolios and G.J. Farooq-Smith and K.T. Flanagan and R.F. {Garcia Ruiz} and W. Gins and others},
    doi = {https://doi.org/10.1016/j.nimb.2019.04.049},
    issn = {0168-583X},
    journal = {Nucl. Instrum. Methods Phys. Res. B},
    pages = {384-389},
    title = {Optimising the Collinear Resonance Ionisation Spectroscopy (CRIS) experiment at CERN-ISOLDE},
    volume = {463},
    year = {2020}
}

@misc{steck_rubidium_2001,
    author = {Steck, D.A.},
    title = {Rubidium 87 D Line Data},
    url = {http://steck.us/alkalidata},
    year = {2001}
}

@article{GINS2018286,
    author = {W. Gins and R.P. {de Groote} and M.L. Bissell and C. {Granados Buitrago} and R. Ferrer and K.M. Lynch and G. Neyens and S. Sels},
    doi = {https://doi.org/10.1016/j.cpc.2017.09.012},
    issn = {0010-4655},
    journal = {Comp. Phys. Comm.},
    pages = {286-294},
    title = {Analysis of counting data: Development of the SATLAS Python package},
    volume = {222},
    year = {2018}
}

@article{duong_first_1987,
    author = {Duong, H.T. and Juncar, P. and Liberman, S. and Mueller, A.C. and Neugart, R. and Otten, E.W. and Peuse, B. and Pinard, J. and Stroke, H.H. and Thibault, C. and others},
    doi = {https://doi.org/10.1209/0295-5075/3/2/008},
    journal = {EPL},
    number = {2},
    pages = {175},
    title = {First observation of the blue optical lines of francium},
    volume = {3},
    year = {1987}
}

@article{sansonetti_spectroscopic_2007,
    author = {Sansonetti, J.E.},
    doi = {https://doi.org/10.1063/1.2719251},
    journal = {J. Phys. Chem. Ref. Data},
    number = {2},
    pages = {497--507},
    title = {Spectroscopic Data for Neutral Francium ({FrI})},
    volume = {36},
    year = {2007}
}

@article{budincevic_laser_2014,
    author = {Budinčević, I and Billowes, J. and Bissell, M.L. and Cocolios, T.E. and De Groote, R. P. and De Schepper, S. and Fedosseev, V.N. and Flanagan, K.T. and Franchoo, S. and Ruiz, R.F. Garcia and others},
    doi = {https://doi.org/10.1103/PhysRevC.90.014317},
    journal = {Phys. Rev. C},
    number = {1},
    pages = {014317},
    title = {Laser spectroscopy of francium isotopes at the borders of the region of reflection asymmetry},
    volume = {90},
    year = {2014}
}

@article{Roberts2023,
    author = {Roberts, B.M. and Fairhall, C.J. and Ginges, J.S.M.},
    doi = {https://doi.org/10.1103/PhysRevA.107.052812},
    journal = {Phys. Rev. A},
    number = {5},
    pages = {052812},
    publisher = {APS},
    title = {Electric-dipole transition amplitudes for atoms and ions with one valence electron},
    volume = {107},
    year = {2023}
}

@article{Tang2017,
    author = {Tang, Y-B. and Lou, B-Q. and Shi, T-Y.},
    doi = {https://doi.org/10.1103/PhysRevA.96.022513},
    journal = {Phys. Rev. A},
    number = {2},
    pages = {022513},
    publisher = {APS},
    title = {Relativistic multireference coupled-cluster theory based on a B-spline basis: Application to atomic francium},
    volume = {96},
    year = {2017}
}

@article{aubin_lifetimes_2004,
    author = {Aubin, S. and Gomez, E. and Orozco, L. A. and Sprouse, G. D.},
    doi = {https://doi.org/10.1103/PhysRevA.70.042504},
    journal = {Phys. Rev. A},
    number = {4},
    pages = {042504},
    title = {Lifetimes of the 9s and 8p levels of atomic francium},
    volume = {70},
    year = {2004}
}

@article{Sahoo_relativistic_2018,
    author = {Sahoo, B.K. and Das, B.P.},
    doi = {https://doi.org/10.1103/PhysRevLett.120.203001},
    issue = {20},
    journal = {Phys. Rev. Lett.},
    month = {May},
    numpages = {6},
    pages = {203001},
    publisher = {American Physical Society},
    title = {Relativistic Normal Coupled-Cluster Theory for Accurate Determination of Electric Dipole Moments of Atoms: First Application to the $^{199}\mathrm{Hg}$ Atom},
    volume = {120},
    year = {2018}
}

@article{yamanaka_probing_2017,
    author = {Yamanaka, N. and Sahoo, B.K. and Yoshinaga, N. and Sato, T. and Asahi, K. and Das, B.P.},
    doi = {https://doi.org/10.1140/epja/i2017-12237-2},
    issn = {1434-601X},
    journal = {EPJA},
    language = {en},
    month = {March},
    number = {3},
    pages = {54},
    shorttitle = {Probing exotic phenomena at the interface of nuclear and particle physics with the electric dipole moments of diamagnetic atoms},
    title = {Probing exotic phenomena at the interface of nuclear and particle physics with the electric dipole moments of diamagnetic atoms: {A} unique window to hadronic and semi-leptonic {CP} violation},
    urldate = {2025-09-16},
    volume = {53},
    year = {2017}
}

@article{gomez_lifetime_2005,
    author = {Gomez, E. and Orozco, {L.A.} and Galvan, A. Perez and Sprouse, {G.D.}},
    doi = {https://doi.org/10.1103/PhysRevA.71.062504},
    journal = {Phys. Rev. A},
    number = {6},
    pages = {062504},
    title = {Lifetime measurement of the 8 s level in francium},
    volume = {71},
    year = {2005}
}

@article{simsarian_lifetime_1998,
    author = {Simsarian, {J.E.} and Orozco, L. A and Sprouse, {G.D.} and Zhao, {W.Z.}},
    doi = {https://doi.org/10.1103/PhysRevA.57.2448},
    journal = {Phys. Rev. A},
    number = {4},
    pages = {2448},
    title = {Lifetime measurements of the 7 p levels of atomic francium},
    volume = {57},
    year = {1998}
}

@article{grossman_lifetime_2000,
    author = {Grossman, J.M. and Fliller, R.P. and Orozco, L.A. and Pearson, M.R. and Sprouse, G.D.},
    doi = {https://doi.org/10.1103/PhysRevA.62.062502},
    journal = {Phys. Rev. A},
    number = {6},
    pages = {062502},
    title = {Lifetime measurements of the 7D levels of atomic francium},
    volume = {62},
    year = {2000}
}

@article{athanasakis-kaklamanakis_radiative_2024,
    author = {Athanasakis-Kaklamanakis, M. and
Wilkins, S. G. and 
Lassègues, P. and 
Lalanne, L. and 
Reilly, J. R. and 
Ahmad, O. and 
Au, M. and 
Bai, S. W. and 
Berbalk, J. 
and others},
    doi = {https://doi.org/10.1103/PhysRevA.110.L010802},
    journal = {Phys. Rev. A},
    number = {1},
    pages = {L010802},
    title = {Radiative lifetime of the {$A \: ^2\Pi_{1/2}$} state in {RaF} with relevance to laser cooling},
    volume = {110},
    year = {2024}
}

@article{Pal__Rel_2007,
    author = {Pal, R. and Safronova, M.S. and Johnson, W.R. and Derevianko, A. and Porsev, S.G.},
    doi = {https://doi.org/10.1103/PhysRevA.75.042515},
    issue = {4},
    journal = {Phys. Rev. A},
    month = {Apr},
    numpages = {11},
    pages = {042515},
    publisher = {American Physical Society},
    title = {Relativistic coupled-cluster single-double method applied to alkali-metal atoms},
    volume = {75},
    year = {2007}
}

@article{safronova_crit_2011,
    author = {Safronova, M.S. and Safronova, U.I.},
    doi = {https://doi.org/10.1103/PhysRevA.83.052508},
    issue = {5},
    journal = {Phys. Rev. A},
    month = {May},
    numpages = {13},
    pages = {052508},
    publisher = {American Physical Society},
    title = {Critically evaluated theoretical energies, lifetimes, hyperfine constants, and multipole polarizabilities in $^{87}\mathrm{Rb}$},
    volume = {83},
    year = {2011}
}

@article{Vernon2019,
    author = {Vernon, A.R. and Billowes, J. and Binnersley, C.L. and Bissell, M.L. and Cocolios, T.E. and Farooq-Smith, G.J. and Flanagan, K.T. and Garcia Ruiz, R.F. and Gins, W. and de Groote, R.P. and others},
    doi = {https://doi.org/10.1016/j.sab.2019.02.001},
    issn = {0584-8547},
    journal = {Spec. Acta B},
    pages = {61--83},
    title = {Simulation of the relative atomic populations of elements $1\leq Z\leq 89$ following charge exchange tested with collinear resonance ionization spectroscopy of indium},
    volume = {153},
    year = {2019}
}

@article{PhysRevA.95.032502,
    author = {de Groote, R.P. and Verlinde, M. and Sonnenschein, V. and Flanagan, K.T. and Moore, I. and Neyens, G.},
    doi = {https://doi.org/10.1103/PhysRevA.95.032502},
    issue = {3},
    journal = {Phys. Rev. A},
    month = {Mar},
    numpages = {10},
    pages = {032502},
    publisher = {American Physical Society},
    title = {Efficient, high-resolution resonance laser ionization spectroscopy using weak transitions to long-lived excited states},
    volume = {95},
    year = {2017}
}

@article{de_groote_use_2015,
    author = {De Groote, R.P. and Budinčević, I. and Billowes, J. and Bissell, M.L. and Cocolios, T.E. and Farooq-Smith, G.J. and Fedosseev, V.N. and Flanagan, K.T. and Franchoo, S. and Ruiz, R.F. and others},
    doi = {https://doi.org/10.1103/PhysRevLett.115.132501},
    journal = {Phys. Rev. Lett.},
    number = {13},
    pages = {132501},
    title = {Use of a continuous wave laser and pockels cell for sensitive high-resolution collinear resonance ionization spectroscopy},
    volume = {115},
    year = {2015}
}

@article{Morton2000_lr,
    author = {Morton, D.C.},
    doi = {https://doi.org/10.1086/317349},
    journal = {Astrophys.\ J.\ Suppl.\ Ser. },
    number = {2},
    pages = {403--436},
    title = {Atomic Data forResonance Absorption Lines. II. Wavelengths Longward of the Lyman Limit for Heavy Elements},
    volume = {130},
    year = {2000}
}

@article{kugler_isolde_2000,
	title = {The {ISOLDE} facility},
	volume = {129},
	issn = {1572-9540},
	doi = {https://doi.org/10.1023/A:1012603025802},
	language = {en},
	number = {1},
	urldate = {2026-07-01},
	journal = {Hyperfine Interactions},
	author = {Kugler, Erich},
	month = dec,
	year = {2000},
	pages = {23--42},
}

@article{vernon_simulation_2019,
	title = {Simulation of the relative atomic populations of elements 1≤\textit{{Z}}≤89 following charge exchange tested with collinear resonance ionization spectroscopy of indium},
	volume = {153},
	issn = {0584-8547},
	doi = {https://doi.org/10.1016/j.sab.2019.02.001},
	urldate = {2024-10-18},
	journal = {Spectrochimica Acta Part B: Atomic Spectroscopy},
	author = {Vernon, A.R. and Billowes, J. and Binnersley, C.L. and Bissell, M.L. and Cocolios, T. E. and Farooq-Smith, G.J. and Flanagan, K.T. and Garcia Ruiz, R.F. and Gins, W. and de Groote, R.P. and Koszorus, A. and Lynch, K.M. and Neyens, G. and Ricketts, C.M. and Wendt, K.D.A. and Wilkins, S.G. and Yang, X.F.},
	month = mar,
	year = {2019},
	pages = {61--83},
}

@article{mattolat_allsolidstate_2009,
	title = {An {All}‐{Solid}‐{State} {High} {Repetiton} {Rate} {Titanium}:{Sapphire} {Laser} {System} {For} {Resonance} {Ionization} {Laser} {Ion} {Sources}},
	volume = {1104},
	issn = {0094-243X},
	shorttitle = {An {All}‐{Solid}‐{State} {High} {Repetiton} {Rate} {Titanium}},
	doi = {https://doi.org/10.1063/1.3115586},
	number = {1},
	urldate = {2026-07-01},
	journal = {AIP Conference Proceedings},
	author = {Mattolat, C. and Rothe, S. and Schwellnus, F. and Gottwald, T. and Raeder, S. and Wendt, K.},
	month = mar,
	year = {2009},
	pages = {114--119},
}

@article{hughes_coupled-cluster_1993,
	title = {The coupled-cluster method with full inclusion of single, double and triple excitations applied to high sectors of the {Fock} space},
	volume = {204},
	issn = {0009-2614},
	doi = {https://doi.org/10.1016/0009-2614(93)90019-W},
	number = {3},
	urldate = {2026-07-01},
	journal = {Chemical Physics Letters},
	author = {Hughes, S. R. and Kaldor, Uzi},
	month = mar,
	year = {1993},
	pages = {339--342},
}

@article{pal_relativistic_2007,
	title = {Relativistic coupled-cluster single-double method applied to alkali-metal atoms},
	volume = {75},
	doi = {https://doi.org/10.1103/PhysRevA.75.042515},
	number = {4},
	urldate = {2026-07-01},
	journal = {Physical Review A},
	publisher = {American Physical Society},
	author = {Pal, Rupsi and Safronova, M. S. and Johnson, W. R. and Derevianko, Andrei and Porsev, Sergey G.},
	month = apr,
	year = {2007},
	pages = {042515},
}

@article{Allegrini_2022,
    author = {Allegrini, Maria and Arimondo, Ennio and Orozco, Luis A.},
    title = {Survey of Hyperfine Structure Measurements in Alkali Atoms},
    journal = {Journal of Physical and Chemical Reference Data},
    volume = {51},
    number = {4},
    pages = {043102},
    year = {2022},
    month = {11},
    issn = {0047-2689},
    doi = {10.1063/5.0098061},
    url = {https://doi.org/10.1063/5.0098061},
    eprint = {https://pubs.aip.org/aip/jpr/article-pdf/doi/10.1063/5.0098061/16597975/043102_1_online.pdf},
}

\end{document}